\documentclass[superscriptaddress,twocolumn,secnumarabic,nofootinbib]{revtex4-1}
\usepackage{amsmath}

\pdfoutput=1

\usepackage{bm}
\usepackage{times}
\usepackage{braket}
\usepackage{color,graphicx,colortbl}
\usepackage[utf8]{inputenc}
\usepackage[T1]{fontenc}
\usepackage{bbold}

\newcommand{\invfb}{{fb$^{-1}$}}

\newcommand{\overbar}[1]{\mkern 1.5mu\overline{\mkern-1.5mu#1\mkern-1.5mu}\mkern 1.5mu}

\usepackage[mathscr]{euscript}

\usepackage[normalem]{ulem}

\definecolor{nicered}{rgb}{0.7,0.1,0.1}
\definecolor{nicegreen}{rgb}{0.1,0.5,0.1}

\usepackage{longtable}

\usepackage{hyperref}
\hypersetup{colorlinks,citecolor=nicegreen,linkcolor=nicered}
\hypersetup{colorlinks=true}

\definecolor{LightCyan}{rgb}{0.8,0.9,0.99}

\begin{document}

{\flushright
{\color{blue}{ \hfill}\\
\color{blue}{ZU-TH 02/20}}\\
\hfill 
}

\title{Lepton Flavor Violation and Dilepton Tails at the LHC}

\author{Andrei Angelescu} \email[]{aangelescu2@unl.edu}
\affiliation{Department of Physics and Astronomy, University of Nebraska-Lincoln, Lincoln, NE, 68588, USA}
\author{Darius A. Faroughy} \email[]{faroughy@physik.uzh.ch }
\affiliation{Physik-Institut, Universität Zürich, CH-8057 Zürich, Switzerland}
\author{Olcyr Sumensari} \email[]{olcyr.sumensari@pd.infn.it}
\affiliation{Dipartimento di Fisica e Astronomia ``G.\ Galilei'', Universit\` a di Padova, Italy}
\affiliation{Istituto Nazionale Fisica Nucleare, Sezione di Padova, I-35131 Padova, Italy}  
 
\begin{abstract}
Starting from a general effective Lagrangian for lepton flavor violation (LFV) in quark-lepton transitions, we derive constraints on the effective coefficients from the high-mass tails of the dilepton processes $pp \to \ell_k \ell_l$ (with $k\neq l$). The current (projected) limits derived in this paper from LHC data with $36~\mathrm{fb}^{-1}$ ($3~\mathrm{ab}^{-1}$) can be applied to generic new physics scenarios, including the ones with scalar, vector and tensor effective operators. For purely left-handed operators, we explicitly compare these LHC constraints with the ones derived from flavor-physics observables, illustrating the complementarity of these different probes. While flavor physics is typically more constraining for quark-flavor violating operators, we find that LHC provides the most stringent limits on several flavor-conserving ones. Furthermore, we show that dilepton tails offer the best probes for charm-quark transitions at current luminosities and that they provide competitive limits for tauonic $b\to d$ transitions at the high-luminosity LHC phase. As a by-product, we also provide general numerical expressions for several low-energy LFV processes, such as the semi-leptonic decays $K\to \pi \ell^{\pm}_k \ell^{\mp}_l$, $B\to \pi \ell^{\pm}_k \ell^{\mp}_l$ and $B\to K^{(\ast)} \ell^{\pm}_k \ell^{\mp}_l$.
\end{abstract}
\maketitle

\section{Introduction}
\label{sec:intro}

Lepton flavor symmetry is accidental in the SM and it is known to be explicitly broken by the nonzero neutrino masses and mixing, as established by neutrino oscillation experiments. Neutrino masses are also responsible for flavor violation in the charged-lepton sector, with unobservable rates suppressed by $(m_\nu/m_W)^4\approx 10^{-48}$. This makes charged Lepton Flavor Violation (LFV) an appealing target for experimental searches beyond the SM (BSM), as its observation would clearly point to the existence of new phenomena. 

From a theoretical perspective, LFV is predicted in various BSM scenarios, such as the ones involving sterile neutrinos~\cite{Petcov:1976ff}, extended Higgs sectors~\cite{Paradisi:2005tk}, $Z^\prime$ bosons~\cite{Langacker:2008yv} and leptoquarks~\cite{Becirevic:2016oho}. Under the assumption of heavy new physics states, the low-energy LFV data can be described by means of an Effective Field Theory (EFT), with the information on the underlying dynamics encoded in effective coefficients that can be probed experimentally.

On the experimental side, there is a rich flavor-physics program dedicated to LFV in both lepton and meson decays. The current sensitivity will be significantly improved  in the coming years by the ongoing effort at the present NA62~\cite{NA62:2017rwk}, LHCb~\cite{Bediaga:2018lhg} and Belle-II~\cite{Kou:2018nap}, as well as at the future Mu2E~\cite{Bartoszek:2014mya}, Mu3E~\cite{Blondel:2013ia} and COMET~\cite{Adamov:2018vin} experiments. While up to date there is no evidence for charged LFV, there are hints of Lepton Flavor Universality Violation (LFUV) in $B$-meson semi-leptonic decays (see e.g.~Ref.~\cite{Bifani:2018zmi} for a recent review), which have attracted a lot of attention in the particle physics community. 
Notably, several BSM resolutions of these discrepancies predict sizeable LFV effects in semi-leptonic operators, see e.g.~\cite{Glashow:2014iga} and references therein. 

In recent years, the large luminosity accumulated at the LHC has offered many opportunities to indirectly test flavor-physics scenarios at high-$p_T$.  In particular, recasts of resonant searches in the invariant mass tails of the $pp\to \ell^-\ell^+$ and $pp\to\ell^\pm\nu_\ell$ processes have been used to derive stringent limits on various new physics models~\cite{Faroughy:2016osc,Greljo:2017vvb,Greljo:2018tzh,Fuentes-Martin:2020lea}. These constraints turn out to be complementary to the ones coming from flavor physics observables and, in particular, they have been useful to identify the viable solutions of the LFUV anomalies observed in $B$-meson decays~\cite{Angelescu:2018tyl,Baker:2019sli}. The main focus of this study is to perform an analogous analysis of the LFV processes $pp \to \ell_k\ell_l$ (with $k \neq l$) at the LHC, which have not been thoroughly explored thus far, and which can also provide complementary information to low-energy observables.

In this paper, we derive constraints on four-fermion LFV operators by using LHC data. To this purpose, we formulate an EFT with generic semileptonic dimension-6 operators and we study their impact onto the LFV dilepton tails at the LHC. Previous phenomenological analyses have considered effective operators with particular Lorentz and/or flavor structures~\cite{Han:2010sa,Cai:2015poa}. We update and extend these analyses by considering the most recent LHC data, as well by accounting for the most general effective operators. Furthermore, for a specific example with left-handed operators, we explicitly compare the high-$p_T$ limits derived in this paper with the ones obtained from low-energy data, by showing their complementarity.

The remainder of the paper is organized as follows. In Sec.~\ref{sec:LHC} we define our setup, we describe the details of our recast of LHC data and derive the corresponding limits. In Sec.~\ref{sec:lowenergy} we derive constraints, by using flavor physics observables, on a specific scenario with purely left-handed operators, which are then compared with the high-$p_T$ limits we have derived in Sec.~\ref{sec:discussion}. Our findings are summarized in Sec.~\ref{sec:conclusion}.

\section{LFV tails at the LHC}
\label{sec:LHC}

\subsection{Framework}
\label{ssec:framework}

We start by defining our framework. We consider the following dimension-6 effective Lagrangian,
\begin{align}
\label{eq:left}
\mathcal{L}_{\mathrm{eff}} & \supset \sum_{\alpha}\sum_{ijkl}\dfrac{C_\alpha^{ijkl}}{v^2}\, \mathcal{O}_{\alpha}^{ijkl}\,,
\end{align}
\noindent where $v=(\sqrt{2} G_f)^{-1/2}$ is the electroweak vacuum expectation value, $\mathcal{O}_\alpha^{ijkl}$ are the semi-leptonic operators collected in Table~\ref{tab:operators} and $C_\alpha^{ijkl}$ are the corresponding effective coefficients. The index $\alpha$ accounts for the possible Lorentz structures, while $\lbrace i,j,k,l\rbrace$ denotes flavor indices. Note that $q_{i,j}$ can be either up or down-type quarks in our notation. Furthermore, dipole operators are not considered in Eq.~\eqref{eq:left} since these are already tightly constrained by radiative LFV decays~\cite{Tanabashi:2018oca}.

Under the assumption of heavy new physics, which we adopt henceforth, the Wilson coefficients in Eq.~\eqref{eq:left} should be matched onto the $SU(3)_c\times SU(2)_L \times U(1)_Y$ invariant basis of dimension-6 operators, as given in the last column of Table~\ref{tab:operators}~\cite{Buchmuller:1985jz,Grzadkowski:2010es}. From this matching, we learn that the vectorial coefficients $C_{V_{XY}}$ (with $X,Y\in \lbrace L, R \rbrace$) can couple to both down and up-type quarks for all possible chirality combinations. On the other hand, if we restrict ourselves to dimension-6 operators, $C_{S_R}$ can only be generated for down-type quarks, while $C_{S_L}$ and $C_T$ only appear for up-type quarks. The complete details of this matching are provided in Appendix~\ref{app:warsaw}.

With the Lagrangian defined above, one can compute the partonic cross-section for $q_i \bar{q}_j \to \ell_k^- \ell_l^+$, with $k\neq l$, at leading order. By neglecting the fermion masses, we can generically express the differential partonic cross-section for this process as 
\begin{align}
\bigg{[}\dfrac{\mathrm{d}\hat{\sigma}}{\mathrm{d}\hat{t}}&\bigg{]}_{ijkl}=\dfrac{(\hat{s}+\hat{t})^2}{48\pi v^4\hat{s}^2}\bigg{\lbrace}\Big[|C_{V_{LL}}|^2+|C_{V_{LR}}|^2+(L\leftrightarrow R)\Big]\nonumber\\
&+\dfrac{\hat{s}^2}{4(\hat{s}+\hat{t})^2}\Big[|C_{S_L}|^2+|C_{S_R}|^2\Big]+\dfrac{4(\hat{s}+2\hat{t})^2}{(\hat{s}+\hat{t})^2}|C_T|^2\nonumber\\&- \dfrac{2\,\hat{s}(\hat{s}+2\hat{t})}{(\hat{s}+\hat{t})^2}\,\mathrm{Re}\left(C_{S_L}\,C_T^\ast\right)\bigg{\rbrace}\,,
\end{align}
\noindent where $\hat{s}$ denotes the partonic energy and $\hat{t}\in(-\hat{s},0)$. After integration, we obtain
\begin{align}
\label{eq:sigma}
\big{[}\hat{\sigma}(\hat s)\big{]}_{ijkl}=\dfrac{\hat s}{144 \pi \, v^4}  \sum_{\alpha\beta} C_\alpha C_\beta^\ast \,{M}_{\alpha\beta}\,,
\end{align}

\noindent where $\alpha,\beta\in \lbrace V_{LL} ,V_{RR},V_{LR},V_{RL},S_{L},S_{R},T    \rbrace$ and $M_{\alpha\beta}$ is a matrix of numeric coefficients. In this equation, chirality-conserving operators should be replaced by
\begin{equation}
C_{V_{X,Y}} \to C_{V_{X,Y}}^{ijkl}\,,
\end{equation}
\noindent with $X,Y\in \lbrace L,R \rbrace$, while the replacement for the chirality-breaking ones reads
\begin{align}
\label{eq:replacement-ST}
\begin{split}
C_{S_X} &\to \sqrt{\big|C_{S_X}^{ijkl}\big|^2+\big|C_{S_X}^{jilk}\big|^2}\,,\\[0.3em]
C_{T} &\to \sqrt{\big|C_{T}^{ijkl}\big|^2+\big|C_{T}^{jilk}\big|^2}\,.
\end{split}
\end{align}
\noindent The terms with inverted flavor indices in Eq.~\eqref{eq:replacement-ST} arise from the Hermitian conjugates in Table~\ref{tab:operators}. Since fermion masses are negligible in this process, the off-diagonal elements of $M$ vanish, so that $M_{\alpha\beta}\equiv \delta_{\alpha\beta} \, M_\alpha$, with
\begin{equation}
\label{eq:M-diag}
M = \Big{(}1,1,1,1,\dfrac{3}{4},\dfrac{3}{4},4\Big{)}\,.
\end{equation}

\begin{table}[t]
\renewcommand{\arraystretch}{1.9}
\centering
\begin{tabular}{|c|c||c|}
\hline 
Eff.~coeff. & Operator & SMEFT \\ \hline\hline
$C_{V_{LL}}^{ijkl}$  & $\big{(}\overline{q}_{Li} \gamma_\mu q_{Lj}\big{)}\big{(}\bar{\ell}_{Lk} \gamma^\mu \ell_{Ll}\big{)}$ & $\mathcal{O}_{lq}^{(1)}$, $\mathcal{O}_{lq}^{(3)}$\\ 
$C_{V_{RR}}^{ijkl}$  & $\big{(}\overline{q}_{Ri} \gamma_\mu q_{Rj}\big{)}\big{(}\bar{\ell}_{Rk} \gamma^\mu \ell_{Rl}\big{)}$ & $\mathcal{O}_{ed}$, $\mathcal{O}_{eu}$ \\
$C_{V_{LR}}^{ijkl}$   & $\big{(}\overline{q}_{Li} \gamma_\mu q_{Lj}\big{)}\big{(}\bar{\ell}_{Rk} \gamma^\mu \ell_{Rl}\big{)}$  & $\mathcal{O}_{qe}$ \\ 
$C_{V_{RL}}^{ijkl}$  & $\big{(}\overline{q}_{Ri} \gamma_\mu q_{Rj}\big{)}\big{(}\bar{\ell}_{Lk} \gamma^\mu \ell_{Ll}\big{)}$ & $\mathcal{O}_{lu}$, $\mathcal{O}_{ld}$\\ 
$C_{S_{R}}^{ijkl}$   & $\big{(}\overline{q}_{Ri} q_{Lj}\big{)}\big{(}\bar{\ell}_{Lk}\ell_{Rl}\big{)}+\mathrm{h.c.}$ & $\mathcal{O}_{ledq}$ \\
$C_{S_{L}}^{ijkl}$    & $\big{(}\overline{q}_{Li} q_{Rj}\big{)}\big{(}\bar{\ell}_{Lk}\ell_{Rl}\big{)}+\mathrm{h.c.}$ & $\mathcal{O}_{lequ}^{(1)}$ \\  
$C_{T}^{ijkl}$ & $\big{(}\overline{q}_{Li} \sigma_{\mu\nu} q_{Rj}\big{)}\big{(}\bar{\ell}_{Lk} \sigma^{\mu\nu} \ell_{Rl}\big{)}+\mathrm{h.c.}$ & $\mathcal{O}_{lequ}^{(3)}$ \\  \hline
\end{tabular}
\caption{ \sl Operators $\mathcal{O}_\alpha$ appearing in Eq.~\eqref{eq:left} and their corresponding operators in the SMEFT (third column). Flavor indices are denoted by $i,j,k,l$, and $q$ stands for either up or down-type quarks in the mass basis. Wilson coefficients are assumed to be real. See Appendix~\ref{app:warsaw} for details.}
\label{tab:operators} 
\end{table}

\begin{figure*}[t!]
  \centering
\includegraphics[width=0.47\textwidth]{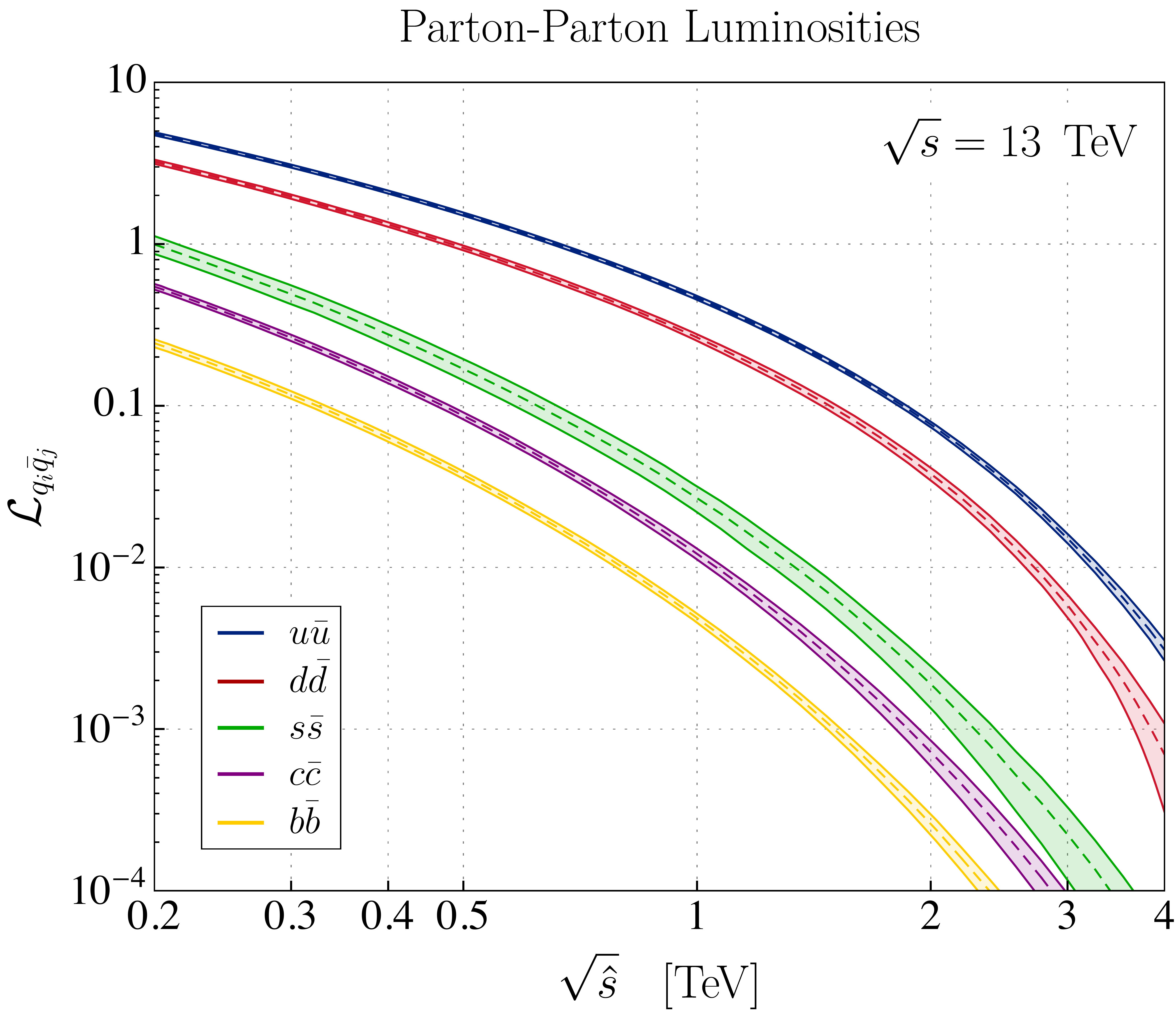}~\qquad~\includegraphics[width=0.47\textwidth]{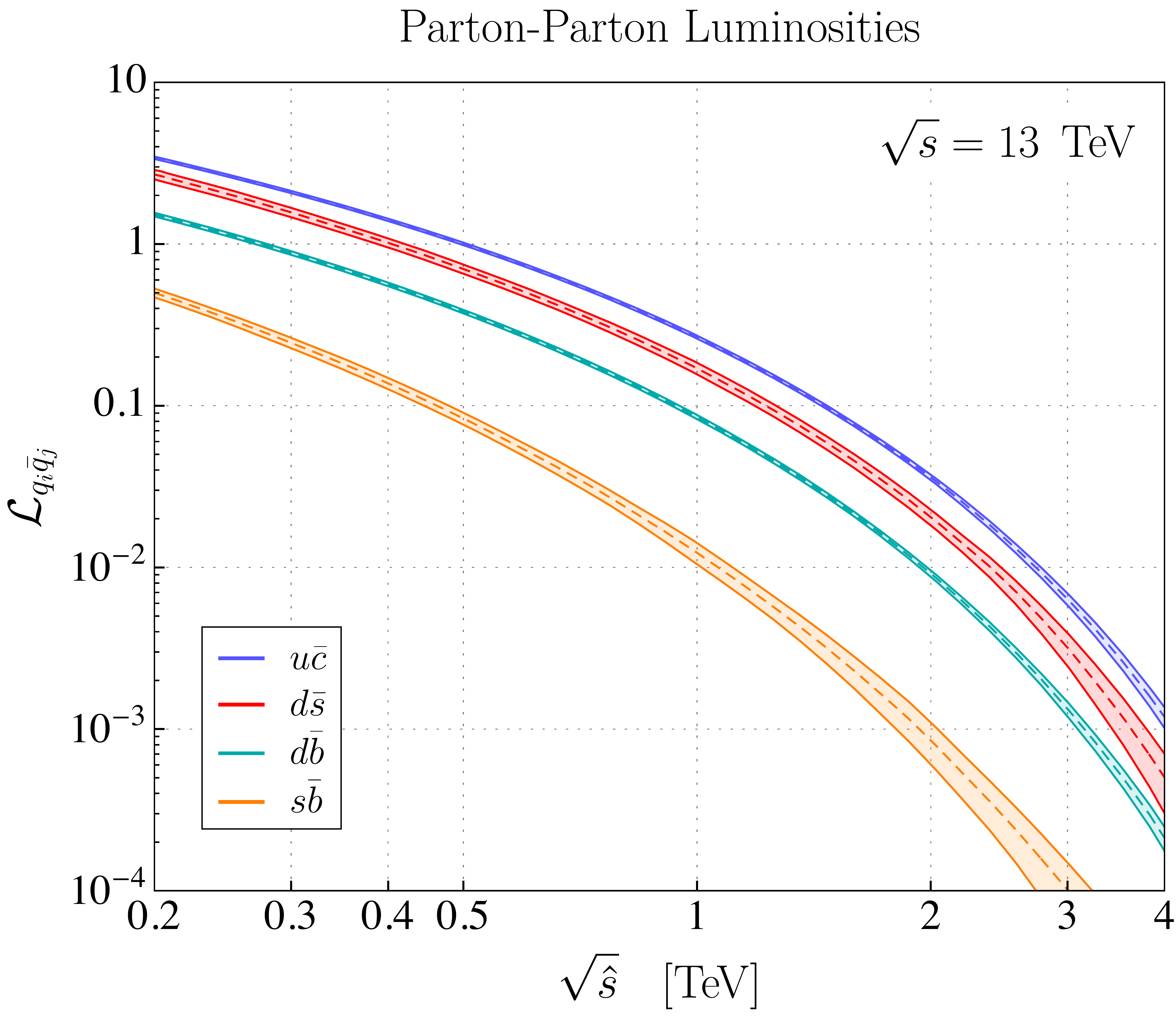}
  \caption{ \sl \small {Parton-parton luminosity functions $\mathcal{L}_{q_i \bar{q}_j}$ (see~Eq.~\eqref{eq:lumidef}) are depicted for quark-flavor conserving and violating processes in the left and right panels, respectively. The PDF set  {\tt PDF4LHC15\_nnlo\_mc} \cite{Butterworth:2015oua,Ball:2014uwa,Harland-Lang:2014zoa,Dulat:2015mca} has been used to extract the central value (dashed lines) and the $1\sigma$ contours (solid envelope).}}
  \label{fig:lumi}
\end{figure*}

\noindent where we use the same ordering of effective coefficients as in Eq.~\eqref{eq:sigma}. The values reported in Eq.~\eqref{eq:M-diag} result from integrating over the full range of angular variables, i.e. $\hat{t}\in(-\hat{s},0)$, which corresponds to the lepton rapidity interval $\eta \in (-\infty,\infty)$. For the recast of LHC searches in Sec.~\ref{sec:LHC}, a rapidity selection cut for final state leptons of $\eta \in (-2.5,2.5)$ introduces an operator dependent angular efficiency ${\epsilon_\alpha}$ not considered in Eq.~\eqref{eq:sigma}. We have explicitly checked that these selection efficiencies are approximately $98\%$, $99\%$, and $96\%$ for the vector, scalar, and tensor operators, respectively, making Eq.~\eqref{eq:sigma} a good approximation. The partonic cross-section should be convoluted with the relevant parton-parton luminosities \cite{Campbell:2006wx}, which in this work we define by the dimensionless functions\footnote{This definition of the parton luminosity functions differs from the one in \cite{Campbell:2006wx} by a multiplicative factor of $\hat s$.}
\begin{align}
\label{eq:lumidef}
\mathcal{L}_{q_i\bar q_j}(\tau) = \tau\int_{\tau}^{1}\dfrac{\mathrm{d}x}{x} \big{[}f_{q_i}(x,\mu_F) f_{\bar{q}_j}(\tau/x,\mu_F)+(q_i \leftrightarrow \bar{q}_j)\big{]}\,, 
\end{align}
where $f_{q_i}$ denotes the quark $q_i$ parton distribution functions (PDF), $\mu_F$ is the factorization scale and $\sqrt{s}$ stands for the proton-proton center-of-mass energy, with $\tau=\hat s/s$. The non-trivial flavor hierarchies of the luminosity functions for different pairs of colliding partons are depicted in Fig.~\ref{fig:lumi} for $\mu_F=\tau s$, where we have used the {\tt PDF4LHC15\_nnlo\_mc} PDF set \cite{Butterworth:2015oua,Ball:2014uwa,Harland-Lang:2014zoa,Dulat:2015mca} and included the $1\sigma$ PDF uncertainties derived by using the MC replica method \cite{Carrazza:2015hva}. The hadronic cross-section is then given by the expression
\begin{equation}
\label{eq:sigma-convoluted}
\sigma(pp\to \ell_k^- \ell_l^+)=\sum_{ij} \int\frac{\mathrm{d}\tau}{\tau}\, \mathcal{L}_{q_i\bar q_j}(\tau) \, \big{[}\hat{\sigma}(\tau s)\big{]}_{ijkl}\,,
\end{equation}
\noindent where $q$ denotes both down and up-type quarks. The summation extends over all quark flavors, with the exception of the top quark which only contributes at one-loop to this process~\cite{Bhattacharya:2018ryy,Cai:2018cog}. Notice that if the partonic cross-section $\hat \sigma$ is a linear function in $\tau$, as it is our case, then the only dependence on $\tau$ of the integrand in Eq.~\eqref{eq:sigma-convoluted} comes from the luminosity functions defined in Eq.~\eqref{eq:lumidef}. 

From Eq.~\eqref{eq:M-diag}, we see that the largest partonic cross-section comes from the tensor operator, which is a factor of 4 larger than the vectorial ones. On the other hand, scalar and vector operators have comparable cross-sections. Given the small differences in the angular efficiencies for these operators, the limits derived on a single operator can be easily translated into others by simply accounting for the numerical factors given in Eq.~\eqref{eq:M-diag}. For this reason, we focus in what follows on a single effective coefficient, which we choose to be $C \equiv C_{V_{LL}}$, with flavor indices defined by
\begin{equation}
\label{eq:eft-simplified}
\mathcal{L}_{\mathrm{eff}} \supset \sum_{ijkl} \dfrac{C_{q_i q_j}^{\ell_k\ell_l}}{v^2} \big{(} \bar{q}_{Li} \gamma_\mu q_{Lj}\big{)} \big{(} \bar{\ell}_{Lk}\gamma^\mu\ell_{Ll}\big{)}\,,
\end{equation}
\noindent where $i,j$ are flavor indices of down ($d,s,b$) or u-type quarks ($u,c$), and $k,l$ of charged leptons ($e,\mu,\tau$), in the mass basis. Hermiticity implies that $\big{(}C^{\ell_k \ell_l}_{q_i q_j}\big{)}^\ast = C^{\ell_l \ell_k}_{q_j q_i}$.  In Sec.~\ref{ssec:summary}, we describe how to apply the high-$p_T$ constraints derived for {the Lagrangian given above} to the most general effective scenario in~Eq.~\eqref{eq:left}.

The relevant observable for probing the LFV operators is the high-mass tail of the invariant mass spectrum $m_{\ell_k\ell_l}$ of the final state dilepton. For instance, for the set of left-handed effective operators defined in Eq.~\eqref{eq:eft-simplified}, this observable is computed from the differential hadronic cross-section (Eq.~\eqref{eq:sigma-convoluted}), which is integrated over a fixed interval $\tau \in [\tau_{\text{min}},\tau_{\text{max}}]$,
\begin{align}
\label{eq:tailobs}
\begin{split}
\big{[}\sigma(pp\to \ell_k^\mp \ell_l^\pm)\big{]}_{\tau_{\text{min}}}^{\tau_{\text{max}}}=&\dfrac{s}{144 \pi \, v^4}\,\sum_{i\leq j}\,\int^{\tau_{\text{max}}}_{\tau_{\text{min}}}\mathrm{d}\tau\,\mathcal{L}_{q_i\bar q_j}(\tau)\\ 
&\qquad\times\,\Big{[}|{C}_{q_i q_j}^{{\ell_k\ell_l}}|^2+|{C}_{q_i q_j}^{{\ell_l\ell_k}}|^2\Big{]}\,,
\end{split}
\end{align}
\noindent where we have used the fact LHC searches do not distinguish the charges of the final lepton states. The integration interval is chosen to map a specific invariant mass window into the tail of the dilepton distribution, far away from the SM resonance poles, and we have summed over the lepton charges, i.e.~$\ell_k^\pm \ell_l^\mp \equiv \ell_k^+\ell_l^- + \ell_k^-\ell_l^+$. The choice of the invariant mass windows should ultimately correspond to the most sensitive mass bins in the experiment. Our recast of LHC data will be detailed in Sec.~\ref{ssec:recast}.

Lastly, we briefly discuss the quark-flavor dependence in Eq.~\eqref{eq:sigma-convoluted}. There are two sources of flavor entering the hadronic cross-section: (i) the underlying flavor structure present in the hard partonic process, which is encoded by the effective coefficients, and (ii) the flavor dependent non-perturbative parton distribution functions (PDF) of the proton. Assuming a large scale separation, these structures factorize at leading order as shown in Eq.~\eqref{eq:sigma-convoluted}. For scenarios with effective coefficients that do not distinguish quark flavor, it is clear from Fig.~\ref{fig:lumi} that the leading contribution to the dilepton tails would come from the partonic process initiated by light quarks. This conclusion is no longer valid if the parton luminosities are weighted by effective coefficients that are hierarchical in quark-flavor space, such as scenarios based on a non-universal $U(2)$ flavor symmetry, for which $b$-quarks can induce the largest contribution~\cite{Greljo:2015mma}. Another scenario often considered is  Minimal Flavor Violating (MFV)~\cite{DAmbrosio:2002vsn}. In this case, the parton luminosity functions $\mathcal{L}_{q_i\bar{q}_j}$ should be scaled with the appropriate CKM factors. In the down-quark sector, the individual contributions to the hadronic cross-section should be weighted as $|V_{ti}V_{tj}^\ast|^2\,\mathcal{L}_{d_i\bar{d}_j}$, for $i\neq j$, suppressing the flavor changing transitions, i.e.~$s\bar{d}$, $b\bar{d}$ and $b \bar{s}$, which become then comparable. 

\subsection{Recast of existing LFV searches}
\label{ssec:recast}

We first implemented the effective Lagrangian~\eqref{eq:eft-simplified} in {\tt FeynRules} \cite{Alloul:2013bka}. After importing the resulting UFO model into {\tt Madgraph5} \cite{Alwall:2014hca}, we simulated statistically significant event samples of $pp \to e^\pm \mu^\mp$, $e^\pm \tau^\mp$, and $\mu^\pm \tau^\mp$ for each combination of initial flavor quarks: $u\bar u$, $d\bar d$, $s\bar s$, $c\bar c$, $b\bar b$, $u\bar c$, $d\bar b$, $d\bar s$, $s\bar b$, as well as their Hermitian conjugates. Each sample was then showered and hadronized using {\tt Pythia8} \cite{Sjostrand:2014zea}. For final state object reconstruction and detector simulation we used the fast simulator {\tt Delphes3} \cite{deFavereau:2013fsa} with parameters tuned to the experimental searches described right below. Jets were clustered with the anti-$k_T$ algorithm with a cone of size $\Delta R=0.4$ using {\tt fastJet} \cite{Cacciari:2011ma}.

\begin{table}[t!]
\renewcommand{\arraystretch}{2.3}
\centering
\begin{tabular}{|c|c c c|}
\hline 
{$\,C_{\mathrm{eff}}\left(\times 10^3\right)\,$} &$e\mu$ &\ $e \tau$\ & $\mu\tau$\\ \hline\hline
$uu$ &\;\;1.0 (0.3)& \;\;2.6 (0.5)& \;\;3.0 (0.7)\ \ \\ 
$dd$ &\;\;1.4 (0.5)& \;\;4.1 (0.9)& \;\;4.5 (1.2)\\ 
$ss$ &\;\;6.5 (2.4)& \;\;21 (5.3)& \;\;22 (6.7)\\ 
$cc$ &\;\;10 (4.0)& \;\;35 (9.5)&\;\;36 (11)\\ 
$bb$ &\;\;18 (6.8)& \;\;59 (17)& \;\;62 (21)\\ \hline\hline
$uc$ &\;\;2.0 (0.7)& \;\;5.8 (1.2)& \;\;6.4 (1.6)\\ 
$ds$ &\;\;2.5 (0.9)& \;\;7.6 (1.7)& \;\;8.2 (2.2)\\
$db$ &\;\;3.9 (1.4)& \;\;12 (2.8)& \;\;13 (3.6)\\ 
$sb$ &\;\;9.9 (3.7)& \;\;34 (9.0)& \;\; 37 (11)\\  \hline  
\end{tabular}
\caption{ \sl Current (projected) LHC (HL-LHC) constraints to $2\sigma$ accuracy on the effective coefficients defined in Eq.~\eqref{eq:eft-simplified} for a luminosity of 36.1 fb$^{-1}$ (3 ab$^{-1}$). Since the LHC searches do not distinguish the final lepton charges, these constraints apply to the combination $\sqrt{|{C}_{q_i q_j}^{\ell_k \ell_l}|^2+|{C}_{q_i q_j}^{\ell_k \ell_l}|^2}$, with the lepton (quark flavor) indices depicted in the columns (rows).}
\label{tab:LHC-limits} 
\end{table}

For our recast, we used the latest ATLAS search of heavy vector resonances decaying into a pair of different flavor leptons, $pp\to Z^\prime\to\ell^{\pm}_1\ell^{\mp}_2$, performed at $\sqrt{s}=13~\mathrm{TeV}$ with 36.1~fb$^{-1}$ of $pp$ collision data~\cite{Aaboud:2018jff}. Their search strategy starts by imposing a basic set of $p_T$ and $\eta$ cuts to the reconstructed leptons in each events, for details see Ref.~\cite{Aaboud:2018jff}. $\tau$-leptons were reconstructed using a $\tau$-tagger based on identifying the visible part of the hadronic $\tau$-lepton ($\tau_h$), i.e. the {\it $\tau$-jet} composed of 1-prong or 3-prong pion tracks. Events with exactly two isolated leptons with different flavors (and arbitrary electric charges) were selected and then categorized into the three non-overlapping signal regions denoted by $e\mu$, $e\tau_h$ and $\mu\tau_h$, each corresponding to one of the three LFV decay channels $Z^\prime\to e^\pm\mu^\mp$, $Z^\prime\to e^\pm\tau_h^\mp$ and $Z^\prime\to \mu^\pm\tau_h^\mp$, respectively. Given that the search focuses on the decay of a heavy resonance, the resulting leptonic pair is expected to fly away back-to back in the azimuthal plane. Hence, in order to reduce the leading backgrounds, the cut $|\Delta\phi_{\ell_1\ell_2}|>2.7$ on the leptonic pair was imposed. For the $e\tau_h$ and $\mu\tau_h$ channels the 4-momentum of the hadronic tau $\tau_h$ was reconstructed by adding the 4-momenta of the $\tau$-jet and the missing transverse energy of the event, which is assumed to come exclusively from $\nu_\tau$ and was taken to be collinear with the $\tau$-jet.  After event selection and categorization of events,  the invariant mass spectra for each channel, $m_{e\mu}$, $m_{e\tau}$ and $m_{\mu\tau}$, is reconstructed bin-by-bin.\\

After imposing the same selection cuts described above on each of the $pp\to\ell_k\ell_l$ simulated samples we binned the data into five invariant mass windows defined by the edges $m_{\ell_k\ell_l}\!\in\!\{300,600,1200,2000,3000\}$~GeV including the overflow bin $m_{\ell_k\ell_l}>3000$~GeV, and extracted the event selection efficiency $\epsilon$ and detector acceptance $\mathcal{A}$. The number of signal events per mass bin at $36.1$~\invfb~was estimated by computing the cross-section using Eq.~\eqref{eq:tailobs} rescaled with the corresponding efficiency factor $\epsilon\mathcal{A}$. A statistical analysis was then performed using as input the estimated background events, the background systematic and statistical uncertainties (added in quadrature) and the observed data provided by the ATLAS collaboration in Table~I of Ref.~\cite{Aaboud:2018jff}.  In our analysis we did not include systematics for the signal process. To set limits on each Wilson coefficient, we combined all five invariant mass bins into a likelihood function based on Poissonian distributions. The 95\% confidence level (CL) upper limits were extracted using the CL$_s$ method \cite{Read:2002hq}  with the {\tt pyhf} package \cite{Heinrich:2019}. For High Luminosity (HL) projections, we repeated the procedure above for a luminosity of $3$~ab$^{-1}$~of data expected at the HL-LHC, assuming that the data scales naively with the luminosity ratio and that all uncertainties scale with the square-root of the luminosity ratio. Although this assumption might seem too optimistic, it is worth emphasizing that higher invariant masses will become accessible at the HL-LHC. Therefore, the leading contribution to the future limits will not come from the data in the invariant mass bins used in our projections, but rather from data populating invariant mass bins deeper in the tails that are currently out of reach and that have a larger signal-to-background ratio. For this reason we consider our projections to be a  rather conservative estimate of the full reach of the HL-LHC.

\subsection{Summary of high-$p_T$ constraints}
\label{ssec:summary}

The constraints we obtain for each individual Wilson coefficient $C_{q_i q_j}^{\ell_k \ell_l}$ defined in Eq.~\eqref{eq:eft-simplified} by using $pp\,(q_i \bar q_j)\to\ell_k \ell_l$ data are given in Table~\ref{tab:LHC-limits}. The quark (lepton) flavor combinations are depicted by the rows (columns). Current LHC constraints have been obtained from $36.1$~\invfb LHC data~\cite{Aaboud:2018jff}, while high-luminosity LHC projections have been estimated at $3~\mathrm{ab}^{-1}$, as described above. 

We explain now how to apply the limits provided in Table~\ref{tab:LHC-limits} to scenarios with more than one effective operator, accounting for operators with general Lorentz and quark-flavor structures. This recast is possible since the different contributions do not interfere, being only weighted by the $M_\alpha$ factors in Eq.~\eqref{eq:sigma-convoluted} and the different parton luminosity functions. If we denote by $\zeta^{ijkl}_{V_{LL}}$ the limits extracted in Table~\ref{tab:LHC-limits} for the effective coefficient ${C}_{q_i q_j}^{\ell_k \ell_l} \equiv {C}_{V_{LL}}^{ijkl}$, then the limits on a scenario with several operators can be expressed in the general form,~\footnote{Note that the selection efficiencies of scalar, vector and tensor operators are expected to be very similar for this recast, as explained below Eq.~\eqref{eq:M-diag}.}
\begin{align}
\label{eq:stat}
\sum_{i\leq j}&\left(\zeta_{V_{LL}}^{ijkl}\right)^{-2}\bigg{\lbrace}\sum_{X,Y}\left[\big{|}C_{V_{XY}}^{ijkl}\big{|}^2+\big{|}C_{V_{XY}}^{ijlk}\big{|}^2\right]\\
&\qquad+\dfrac{3}{4}\sum_{X}\left[\big{|}C_{S_{X}}^{ijkl}\big{|}^2+\big{|}C_{S_{X}}^{ijlk}\big{|}^2+(i\leftrightarrow j)\right]\nonumber\\
&\qquad+4\, \left[\big{|}C_{T}^{ijkl}\big{|}^2+\big{|}C_{T}^{ijlk}\big{|}^2+(i\leftrightarrow j)\right]\bigg{\rbrace}\leq 1\,,\nonumber
\end{align}

\noindent where $X,Y=L,R$ and the summation extends over both down and up-type quarks in the mass basis. Lepton flavor indices are fixed since they are constrained by different LHC searches. The numeric pre-factors for scalar and tensor operators correspond to the coefficients $M_{S_X}$ and $M_T$ defined in Eq.~\eqref{eq:M-diag}. The coefficients $C_\alpha^{ijkl}$ appearing implicitly in Eq.~\eqref{eq:stat} can then be explicitly matched onto the SMEFT basis, as described in Appendix~\ref{app:warsaw}. In particular, for operators involving quark doublets, one should account for the flavor mixing induced by the CKM matrix, which can be relevant for certain flavor ansatz.

\section{Low-energy limits}
\label{sec:lowenergy}

In this Section we compare the LHC bounds derived above with the ones obtained from flavor-physics observables at tree-level. The complementarity of both approaches will be illustrated for the purely left-handed operators defined in Eq.~\eqref{eq:eft-simplified}, since LHC and flavor experiments can provide competitive bounds in this case. For these operators, QCD running of the Wilson coefficients is forbidden by the Ward identities, while electroweak and QED running effects are small, allowing for a direct comparison between the two approaches.

There are four types of processes which are relevant for our study: (i) $\mu\to e$ conversion in nuclei, (ii) flavor-changing neutral current (FCNC) decays of mesons, (iii) quarkonium decays, and (iv) hadronic $\tau$ decays. The most up-to-date experimental limits, on which we rely for the analysis in this Section, are listed in Table~\ref{tab:exp}. In the following, we provide the expressions for each of these observables and derive the relevant 2$\sigma$ constraints from existing data.~\footnote{In Appendix~\ref{app:flavor}, we provide the needed theoretical inputs for the most general EFT setup, including scalar and tensor interactions.} 

\

\begin{table}[!t]
\centering
  \renewcommand{\arraystretch}{1.55} 
\begin{tabular}{|c|ccc|}
\hline
\multicolumn{4}{|c|}{Flavor physics limits} \\ \hline\hline
Decay mode & Exp.~limit & Future prospects & Ref.\\
\hline\hline

$K_L\to \mu^\mp e^\pm$	& \;$6.1 \times 10^{-12}$	& -- &	 \cite{Tanabashi:2018oca} \\

$K^+\to \pi^+\mu^+ e^-$	& \;$1.7 \times 10^{-11}$	& $\approx 10^{-12}$ &	 \cite{Tanabashi:2018oca} \\

$\phi\to \mu^\pm e^\mp$	& \;$ 2.6 \times 10^{-6}$ & -- &	\cite{Tanabashi:2018oca}  \\

$D\to \mu^\pm e^\mp$	& \;$1.6\times 10^{-8}$ & -- &	\cite{Tanabashi:2018oca}  \\

$J/\psi \to \mu^\pm e^\mp$	& \;$2.1\times 10^{-7}$	 & -- &	\cite{Tanabashi:2018oca}  \\

$B_d\to \mu^\mp e^\pm$	& \;$1.3\times 10^{-9}$	& $\approx 2\times 10^{-10}$ &	\cite{Aaij:2017cza}  \\

$B^+\to \pi^+ \mu^\mp e^\pm $	& \;$2.2\times 10^{-7}$ & -- &	\cite{Tanabashi:2018oca}  \\

$B_s\to \mu^\mp e^\pm$	& \;$6.3\times 10^{-9}$	& $\approx 8\times 10^{-10}$ &	\cite{Aaij:2017cza}  \\

$B^+\to K^+ \mu^+ e^-$ & \;$8.8 \times 10^{-9}$	& --	&	\cite{Aaij:2019nmj} \\

$B^0\to K^\ast \mu^\mp e^\pm$ & \;$2.3 \times 10^{-7}$ & -- &\cite{Tanabashi:2018oca}	\\ 

\hline \hline

$\tau\to e \rho$	& \;$ 2.3 \times 10^{-8} $	& $\approx 5\times 10^{-10}$ & \cite{Tanabashi:2018oca} 	\\

$\tau \to e K^\ast $	& \;$ 4.2 \times 10^{-8} $	& $\approx  7\times 10^{-10}$ & \cite{Tanabashi:2018oca} 	\\

$\tau \to e \phi $	& \;$ 4.0 \times 10^{-8} $	& $\approx 7\times 10^{-10}$ & \cite{Tanabashi:2018oca} 	\\

$J/ \psi \to \tau^\pm e^\mp $	& \;$ 1.1 \times 10^{-5} $	& -- & \cite{Tanabashi:2018oca} 	\\

$B_d\to \tau^\pm e^\mp$	& \;$3.6\times 10^{-5}$	& $\approx 1.6\times 10^{-5}$ & \cite{Tanabashi:2018oca} 	\\ 

$B^+ \to \pi^+ \tau^\pm e^\mp$	& \;$9.7\times 10^{-5}$	 & -- & \cite{Tanabashi:2018oca} 	\\ 

$B^+\to K^+ \tau^\pm e^\mp$ & \;$3.9 \times 10^{-5}$ &	$\approx 2.1 \times 10^{-6}$& \cite{Tanabashi:2018oca}	\\ 

$ \Upsilon(3S) \to \tau^\pm e^\mp $	& \;$ 5.4 \times 10^{-6} $ & -- & \cite{Tanabashi:2018oca} 	\\

\hline\hline

$\tau\to \mu \rho$	& \;$ 1.6 \times 10^{-8} $	& $\approx 3\times 10^{-10}$ & \cite{Tanabashi:2018oca} 	\\

$\tau\to \mu K^\ast$	& \;$ 7.7 \times 10^{-8} $	& $\approx 10^{-9}$ & \cite{Tanabashi:2018oca} 	\\

$\tau\to \mu \phi$	& \;$ 1.1 \times 10^{-7} $	& $\approx 2\times 10^{-9}$ & \cite{Tanabashi:2018oca} 	\\

$ J/\psi \to \tau^\pm \mu^\mp $	& \; $ 2.6 \times 10^{-6} $ & -- & \cite{Tanabashi:2018oca} 	\\

$B_d\to \tau^\pm \mu^\mp$	& \;$1.4\times 10^{-5}$ & $\approx 1.3\times 10^{-5}$ &	\cite{Aaij:2019okb}  \\ 

$B^+ \to \pi^+ \tau^\pm \mu^\mp$	& \;$ 9.4 \times 10^{-5}$ & -- & \cite{Tanabashi:2018oca} 	\\ 

$B_s\to \tau^\pm \mu^\mp$	& \;$4.2\times 10^{-5}$ & -- &	\cite{Aaij:2019okb} \\ 

$B^+\to K^+ \tau^\pm \mu^\mp$ & \;$ 6.2 \times 10^{-5}$ & $\approx 3.3 \times 10^{-6}$	& \cite{Tanabashi:2018oca}	\\

$ \Upsilon(3S) \to \tau^\pm \mu^\mp $	& \;$ 4.0 \times 10^{-6} $	& -- & \cite{Tanabashi:2018oca} 	\\

\hline
\end{tabular}
\caption{\em \small Most relevant experimental limits at 95\% CL on LFV $\tau$ and leptonic meson decays~\cite{Tanabashi:2018oca} and future prospects for NA62~\cite{Petrov:2017wza}, LHCb~\cite{Bediaga:2018lhg,Borsato:2018tcz} and Belle-II~\cite{Kou:2018nap,Cerri:2018ypt}. Limits available in the literature only at 90\% CL have been rescaled to 95\% CL following Ref.~\cite{Calibbi:2017uvl}. }  
\label{tab:exp}
\end{table}

\subsection{$\mu\to e$ conversion in nuclei}
\label{ssec:mueN}

The strongest bounds on four-fermion operators involving $e \mu$ and first generation quarks come from considering $\mu\to e$ conversion in nuclei. For a nucleus $N$ with atomic number $Z$ and mass number $A$, the expression for the spin-independent conversion rate reads~\cite{Kuno:1999jp},
\begin{align}
{\mathcal B}(\mu \to e, N)_\mathrm{SI} &\simeq \frac{\alpha^3 G_F^2 m_\mu^5 Z_{\rm eff}^{4} F_p^2}{8 \pi^2 Z \, \Gamma_{\rm capt}}\\
&\times\left | (A+Z) C_{uu}^{\mu e} + (2A-Z) C_{dd}^{\mu e} \right |^2 \,,\nonumber
\label{eq:mu-e-conversion}
\end{align}

\noindent with $Z_{\rm eff}$ the effective nuclear electric charge, $F_p$ the nuclear matrix element, and $\Gamma_{\rm capt}$ the muon capture rate. The best current limit on this process comes from measurements performed on ($^{197}_{79}\mathrm{Au}$) nuclei at the SINDRUM-II experiment, and reads ${\rm CR}(\mu \to e, {\rm Au}) < 9.1 \times 10^{-13} $~\cite{Bertl:2006up} at 95\% CL. Using the values for gold nuclei from Ref.~\cite{Kitano:2002mt}, namely $Z_{\rm eff} \simeq 33.5$, $F_p \simeq 0.16$, and $\Gamma_{\rm capt} \simeq 8.6 \times 10^{-18}$~GeV, and considering a single Wilson coefficient at a time, we find the following limits,
\begin{equation}
\left| C_{uu}^{\mu e} \right| < 1.7 \times 10^{-7}\,, \qquad \left| C_{dd}^{\mu e} \right| < 1.5 \times 10^{-7}\,.
\label{eq:mu-e-Au-bound}
\end{equation}

\noindent These bounds are going to be improved by the future experiments COMET~\cite{Adamov:2018vin} and MU2E~\cite{Bartoszek:2014mya} which will use $^{27}_{13}\mathrm{Al}$ targets. For instance, the projected limit from the COMET experiment is ${\rm CR}(\mu \to e, {\rm Al}) \lesssim 10^{-16}$, which will improve the limits in Eq.~\eqref{eq:mu-e-Au-bound} by two orders of magnitude. Such improvement will also open the possibility to probe spin-dependent contributions, such as the one induced for axial-vector operators, which are not coherently enhanced. An interesting example is the effective coefficient $C_{ss}^{\mu e}$, which only contributes via the axial current, since the conservation of the vector current implies that $\langle N| \bar{s}\gamma^\mu s | N\rangle=0$, with $N$ denoting a nucleon. In this case, by using the theoretical inputs provided in Ref.~\cite{Cirigliano:2017azj} for $^{27}_{13}\mathrm{Al}$ targets, and by neglecting the spin-independent contributions, we estimate that the future sensitivity on $C_{ss}^{\mu e}$ will be of $\mathcal{O}(10^{-6})$.

We also note that $C_{uu}^{\mu e}$ and $C_{dd}^{\mu e}$ can be constrained by limits on the LFV decay $\pi^0 \to \mu e$. As we have checked, these limits are orders of magnitude weaker than the ones from $\mu\to e$ conversion, mainly due to the very short lifetime of $\pi^0$.

\subsection{FCNC meson decays}

We consider next quark flavor violating decays of mesons. The simplest observables one can consider are leptonic decays of pseudoscalar mesons, such as $B_s \to \ell_k\ell_l$, with $k>l$. By using the effective Lagrangian~\eqref{eq:eft-simplified}, one can show that
\begin{align}
\begin{split}
\mathcal{B}(B_s \to \ell_k^{\pm} \ell_l^{\mp}) &= \left(|{C}_{bs}^{\ell_k\ell_l}|^2+|{C}_{bs}^{\ell_l\ell_k}|^2\right) \\
&\quad\times\dfrac{f_{B_s}^2 m_{B_s} m_{\ell_k}^2}{64 \pi \Gamma_{B_s}v^4} \bigg{(}1-\dfrac{m_{\ell_k}^2}{m_{B_s}^2}\bigg{)}^2\,,
\label{eq:BR_meson_FCNC}
\end{split}
\end{align}
where we have used $m_{\ell_k}\gg m_{\ell_l}$. In this equation, $f_{B_s}=224(5)$~MeV is the $B_s$-meson decay constant~\cite{Aoki:2016frl}, $\lambda(a^2,b^2,c^2) \equiv (a^2-(b-c)^2)(a^2-(b+c)^2)$ and we have summed over the lepton charges, i.e.~$\ell_1^\pm\ell_2^\mp \equiv \ell_1^+\ell_2^- +  \ell_1^-\ell_2^+$. Expressions for other pseudoscalar meson decays can be obtained by making the suitable replacements. 

Relevant constraints can also be obtained from semi-leptonic decays $P\to P^\prime \ell_1^\pm \ell_2^\mp$, with $P,P^\prime$ being pseudoscalar mesons. The branching ratio expressions can be found Ref.~\cite{Becirevic:2016zri} for the $b\to s\ell_1\ell_2$ transition, which can be easily adapted to the other transitions. We provide the needed expressions and numerical inputs for the most relevant decay modes in Appendix~\ref{app:flavor}. We discuss now each of the relevant observables:

\

\paragraph*{$\bullet$ $\mathbf{s\to d}$:} Contributions from new physics to the transition $s\to d e \mu$ are constrained by the stringent experimental limits listed in Table~\ref{tab:exp}. The most constraining bound is obtained from $K_L\to \mu^\pm e^\mp$ and reads
\begin{align}
|C_{sd}^{e\mu} + C_{sd}^{\mu e}| < 7 \times 10^{-7} \,.
\end{align}

\noindent where we have used $f_K=155.7(0.3)$~\cite{Aoki:2019cca}. Complementary constraints can be extracted from the experimental limits on the $K^+\to\pi^+ e^-\mu^+$ and $K^+\to\pi^+ e^+\mu^-$ decay, cf.~Table~\ref{tab:exp}, which provide separate limits on these effective coefficients, 
\begin{align}
|C_{sd}^{e\mu}|<7\times 10^{-5}\,,\qquad | C_{sd}^{\mu e}| < 5 \times 10^{-5} \,.
\end{align}

\noindent Prospects of improving these limits at LHCb have been recently discussed in Ref.~\cite{Borsato:2018tcz}. 

\

\paragraph*{$\bullet$ $\mathbf{b\to d}$:} Another quark-level transition which is being tested experimentally is the $b\to d \ell_k \ell_l$, with $\ell_{k,l} = e, \mu, \tau$. The relevant decays for this mode are the leptonic $B^0 \to \ell_k \ell_l$ and semi-leptonic $B \to \pi \ell_k \ell_l$ decays. Using the corresponding limits from Table~\ref{tab:exp} and the form factors available from Ref.~\cite{Aoki:2019cca}, we derive the following bounds:

\begin{align}
\sqrt{|C_{db}^{e\mu}|^2+|C_{db}^{\mu e}|^2} &< 3 \times 10^{-4} \,,\\[0.4em]
\sqrt{|C_{db}^{e\tau}|^2+|C_{db}^{\tau e}|^2} &< 5 \times 10^{-3} \,,\\[0.4em]
\sqrt{|C_{db}^{\mu\tau}|^2+|C_{db}^{\tau\mu}|^2} &< 3 \times 10^{-3}  \,.
\end{align}

\

\paragraph*{$\bullet$ $\mathbf{b\to s}$:} Several limits are available for the transition $b\to s \ell_k \ell_l$, with $\ell_{k,l}=e,\mu,\tau$~\cite{Tanabashi:2018oca}, the most constrained modes being the ones with electrons and muons. For the operators we consider, the most constraining limits come from the recent LHCb searches for $\mathcal{B}(B \to K \mu^+ e^-)$ and $\mathcal{B}(B \to K \mu^- e^+)$~\cite{Aaij:2019nmj}. These results independently constrain the Wilson coefficients we consider,
\begin{equation}
|C_{sb}^{e\mu}| < 5 \times 10^{-5} , \qquad |C_{sb}^{\mu e}| < 5 \times 10^{-5} \,.
\end{equation}

\noindent The decay channels with $\tau$'s in the final state face weaker limits. Using the results from Table~\ref{tab:exp}, we obtain
\begin{align}
\sqrt{|C_{sb}^{e\tau}|^2+|C_{sb}^{\tau e}|^2}< 5 \times 10^{-3} \,, \\[0.4em]
\sqrt{|C_{sb}^{\mu\tau}|^2+|C_{sb}^{\tau\mu}|^2}< 5 \times 10^{-3}  \,.
\end{align}

\

\paragraph*{$\bullet$ $\mathbf{c\to u}$:} Finally, let us comment on constraints from $D$-meson decays. In this case, one cannot directly determine the ${\mu\tau}$ coefficient, since the decays $D^0 \to \tau \mu$ and $\tau \to \mu D^0$ are kinematically forbidden. Experimental limits are only available for $\mathcal{B}(D^0\to e^\pm \mu^\mp)$, from which we derive that
\begin{align}
\sqrt{|C_{uc}^{e\mu}|^2+|C_{uc}^{\mu e}|^2}< 5\times 10^{-3} \,.
\end{align}

\noindent Note that limits on semi-leptonic decays $D\to \pi e^{\pm} \mu^{\mp}$ are less constraining than the leptonic ones due to the still weaker experimental sensitivity~\cite{Tanabashi:2018oca}. 

\

\subsection{Quarkonium decays}

The second class of flavor processes we consider are quarkonium decays into leptons. Measurements of such decays represent the only possibility to directly constrain quark-flavor conserving effective coefficients at low-energies. For instance, the decays $\Upsilon\to {\ell_1}^\pm {\ell_2}^\mp$ are induced at tree-level by the operators in Eq.~\eqref{eq:eft-simplified}, giving 
\begin{align*}
\label{eq:BR-quarkonia}
\begin{split}
\mathcal{B}(\Upsilon\to \ell_k^\pm \ell_l^\mp) &= \dfrac{|{C}_{bb}^{\ell_k\ell_l}|^2}{v^4} \dfrac{f_\Upsilon^2 m_\Upsilon^3}{ 24 \pi \Gamma_{\Upsilon}} \left(1-\dfrac{3 m_{\ell_k}^2}{2 m_\Upsilon^2}+\dfrac{m_{\ell_k}^6}{2 m_\Upsilon^6}\right)\,,
\end{split}
\end{align*}

\noindent where we have assumed $m_{\ell_k}\gg m_{\ell_l}$ and $f_{\Upsilon}=649(31)$~MeV is the relevant decay constant~\cite{Colquhoun:2014ica}. Due to Hermiticity, we have $|C_{bb}^{e\mu}|= |C_{bb}^{\mu e}|$. Expressions for the other quarkonium decays can be obtained after making the necessary adjustments. We shall now determine the constraints on new physics from the available experimental results for each transition:

\

\paragraph*{$\bullet$ $\mathbf{s\,\bar{s}}$:} The only kinematically allowed decay of the $\phi$-meson is $\phi\to \mu e$. The experimental limit on  $\mathcal{B}(\phi\to \mu^\pm e^\mp)$ from Table~\ref{tab:exp} can be translated into the bound
\begin{align}
|{C}_{ss}^{e\mu}|< 2\times 10^2\,,
\end{align}

\noindent which is considerably weaker than the limits derived from FCNC decays. Much more stringent limits will be available in the future via spin-dependent $\mu\to e$ conversion in light nuclei, see discussion in Sec.~\ref{ssec:mueN}.

\

\begin{figure*}[t!]
  \centering
    \includegraphics[width=0.3\textwidth]{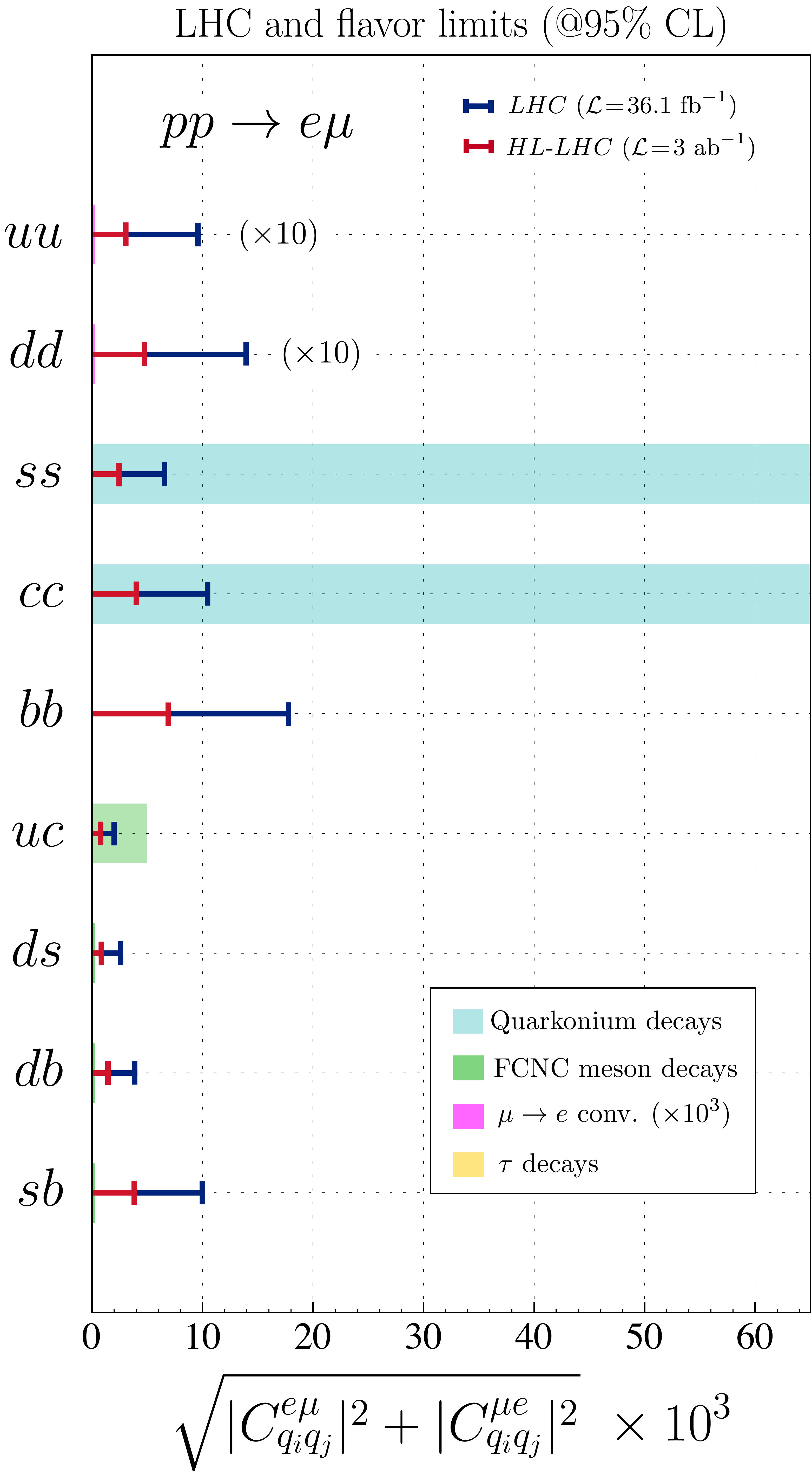}\hspace{5pt}  \includegraphics[width=0.3\textwidth]{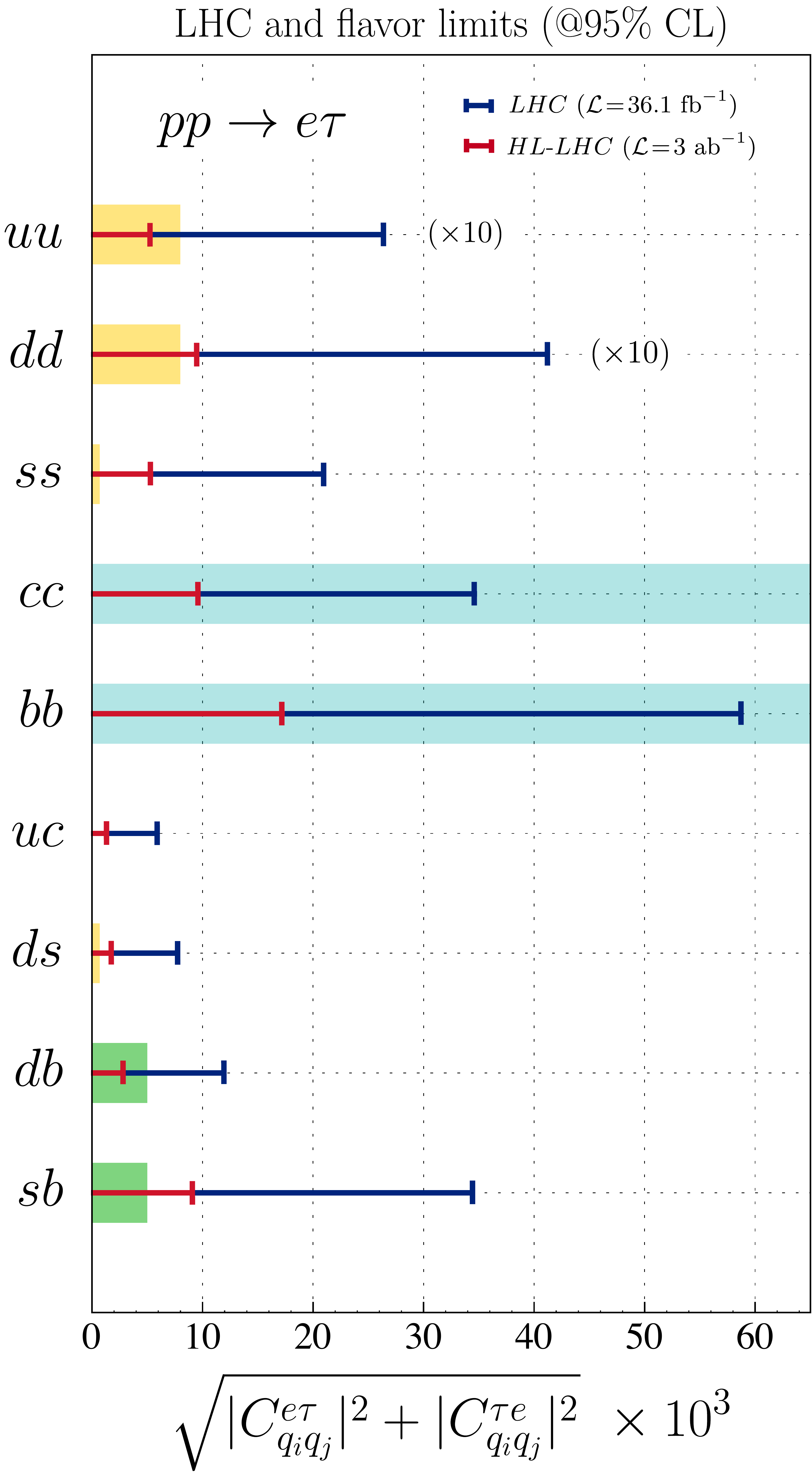}\hspace{5pt} \includegraphics[width=0.3\textwidth]{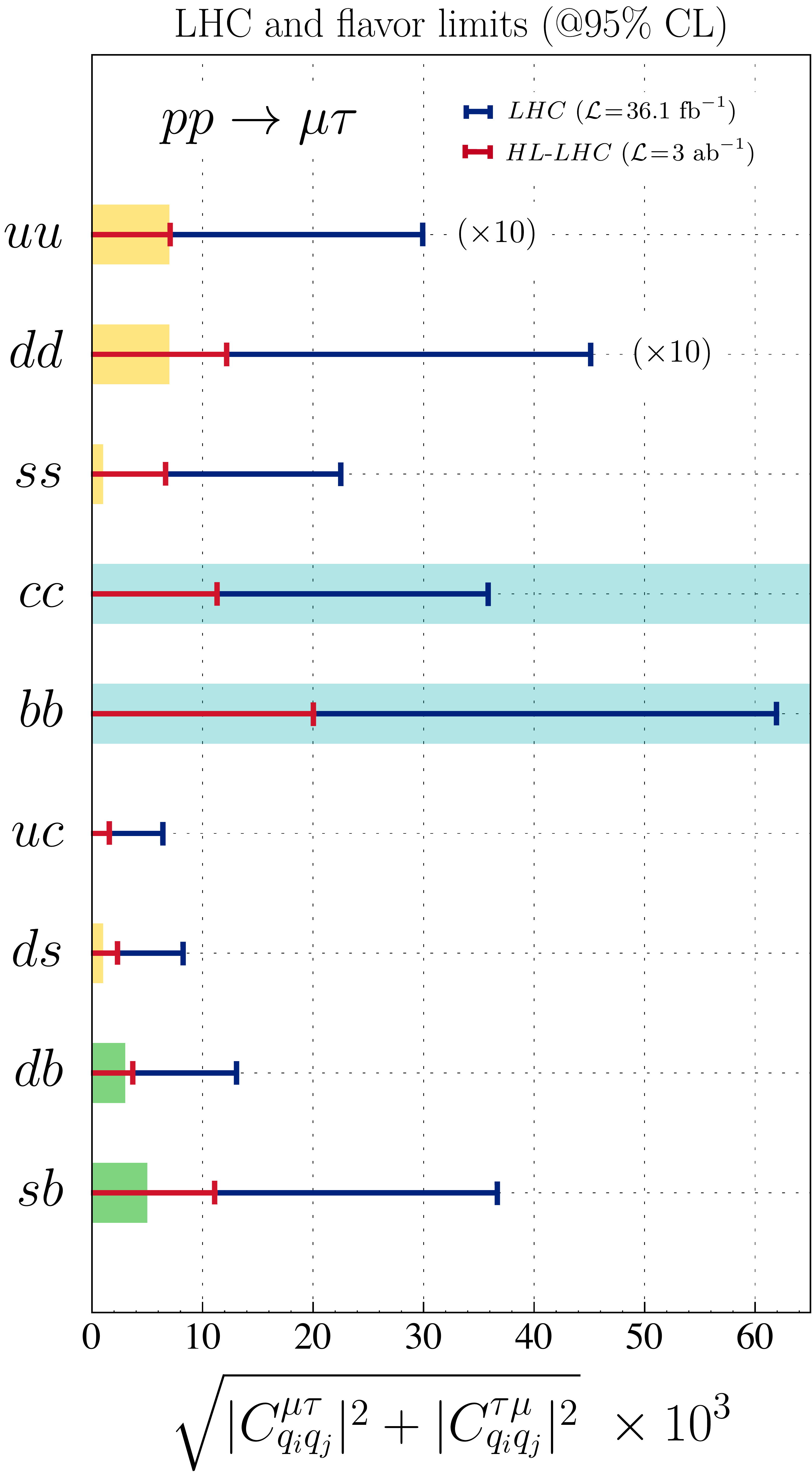}
  \caption{ \sl \small Limits derived from high-$p_T$ LFV dilepton tails on the coefficients $\sqrt{|{C}_{q_iq_j}^{\ell_k\ell_l}|^2+|{C}_{q_iq_j}^{\ell_l\ell_k}|^2}$ by using $13$~TeV ATLAS searches~\cite{Aaboud:2018jff} into the $e\mu$ channel (left panel), the $e\tau$ channel (middle panel) and the $\mu\tau$ channel (right panel). For comparison, we show the limits obtained by the flavor physics observables, namely quarkonium decays (cyan), $\mu N\to e N$ (magenta), FCNC meson decays (green) and LFV $\tau$-decays (yellow). The LHC and flavor results for $uu,dd\to e\mu,e\tau,\mu\tau$ have been rescaled by an additional factor of $\times 10$ for visibility. The limits from $\mu N\to e N$ have been rescaled by a factor of $\times 10^{3}$ to become visible. }
  \label{fig:EFT-comparison}
\end{figure*}

\paragraph*{$\bullet$ $\mathbf{c\,\bar{c}}$:} $J/\psi$ is heavy enough to produce all possible LFV final states. Using the relevant experimental bounds and the decay constant $f_{J/\psi}=418(9)$~MeV~\cite{Becirevic:2012dc}, we are able to determine
\begin{align}
|{C}_{cc}^{e\mu}|< 1.1\,,\quad\;\; |{C}_{cc}^{e\tau}|< 10 \,,\quad\;\; |{C}_{cc}^{\mu\tau}|< 5 \; .
\end{align}

\noindent These limits are considerably weaker than the ones derived above from FCNC decays.

\

\paragraph*{$\bullet$ $\mathbf{b\, \bar{b}}$:} Finally, we discuss LFV decays of $\Upsilon(nS)$ mesons, with experimental bounds only available for $e\tau$ and $\mu\tau$ final states. The most stringent limits on the relevant Wilson coefficients come from $\Upsilon(3S)$ decays. By using the decay constant $f_{\Upsilon(3S)}=539(84)$~MeV~\cite{Lewis:2012ir}, we obtain
\begin{align}
|{C}_{bb}^{e\tau}|< 0.3 \,,\qquad |{C}_{bb}^{\mu\tau}|< 0.3 \,.
\end{align}

\noindent Although these limits are more constraining than the ones derived from $J/\psi$ decays, they remain once again much weaker than the ones coming from FCNC decays. This can be understood from the fact that vector quarkonia have a total width which is orders of magnitude larger than the ones of kaons, $D$ and $B$-mesons. For this reason, such a large width suppresses the branching ratio, making these observables much less sensitive to new physics.

\subsection{$\tau$-lepton decays}

Finally, we turn our attention to LFV hadronic $\tau$ decays. Experimental limits on such processes can constrain the Wilson coefficients $C_{ij}^{\ell \tau}$ ($\ell=e,\mu$), with $i,j=d,s$. Particularly efficient constraints on new physics comes from the decays $\tau\to \phi\, \ell_l$, which are described by,
\begin{align*}
\mathcal{B}(\tau^-\to\phi\,\ell_l^-) =  \dfrac{|{C}_{ss}^{l\tau}|^2}{v^4} \dfrac{f_\phi^2 m_\tau^3}{128 \pi \Gamma_\tau}\left(1-\dfrac{3 m_{V}^2}{m_\tau^2}+2\dfrac{m_{V}^6}{ m_\tau^6}\right)\,,
\end{align*}
from which we derive the following bounds,
\begin{align}
|C_{ss}^{e\tau}|< 7 \times 10^{-4} \,,\qquad|C_{ss}^{\mu\tau}|< 1 \times 10^{-3} \,.
\end{align}

\noindent Similarly, one can use the limits on $\mathcal{B}(\tau \to e\rho)$ and $\mathcal{B}(\tau \to\mu\rho)$ from Table~\ref{tab:exp} to obtain
\begin{align}
\label{eq:eqEx1}
\begin{split}
|C_{uu}^{e\tau} - C_{dd}^{e\tau}| &< 8 \times 10^{-4} \,,  \\[0.4em] |C_{uu}^{\mu\tau} - C_{dd}^{\mu\tau}| &< 7 \times 10^{-4} \,.
\end{split}
\end{align}

\noindent We have checked that analogous limits from $\tau \to \ell \pi$ and $\tau \to \ell \omega$ are less constraining than the ones derived above, see~also~Ref.~\cite{Bordone:2017anc}. Nevertheless, we quote below the limits coming from $\tau \to \ell \omega$, as they probe a different linear combination of Wilson coefficients compared to $\tau \to \ell \rho$. These limits read
\begin{align}
\label{eq:eqEx2}
\begin{split}
|C_{uu}^{e\tau} + C_{dd}^{e\tau}| &< 2 \times 10^{-3} \,,  \\[0.4em] |C_{uu}^{\mu\tau} + C_{dd}^{\mu\tau}| &< 2 \times 10^{-3} \,.
\end{split}
\end{align}
Finally, as there are no experimental bounds on the $\tau \to \ell K_L$ decay, we use the existing limits on $\tau \to \ell K_S$, also listed in Table~\ref{tab:exp}. We find the following constraints:
\begin{align}
\label{eq:eqEx3}
\begin{split}
|C_{ds}^{e\tau} - C_{ds}^{\tau e}| &< 10^{-3} \,, \\[0.3em] |C_{ds}^{\mu\tau} - C_{ds}^{\tau\mu}| &< 10^{-3} \,. 
\end{split}
\end{align}
Note that for scenarios predicting $C_{ds}^{e\tau}= \left(C_{ds}^{\tau e}\right)^{\ast}$, the contributions to $\mathcal{B}(\tau \to \ell K_S)$ would be proportional to the imaginary part of $C_{ds}^{e\tau}$, which we assume to be zero in this study. In this case, an alternative would be to consider bounds on $\mathcal{B}(\tau \to \ell K^{\ast})$ and $\mathcal{B}(\tau \to \ell \overbar{K^{\ast}})$ decays, from which we derive the following bounds, by using the decay constant reported in Ref.~\cite{Allton:2008pn},
\begin{align}
\begin{split}
|C_{ds}^{e\tau}| &< 7 \times 10^{-4} \, , \qquad |C_{ds}^{\tau e}| < 7 \times 10^{-4}\,, \\[0.3em]
|C_{ds}^{\mu\tau}| &< 10^{-3} \, , \qquad \quad \;\;\: |C_{ds}^{\tau \mu}| <  10^{-3} \, .
\end{split}
\end{align}

\section{Results and Discussion}
\label{sec:discussion}

We now compare the constraints on left-handed operators derived in Sec.~\ref{sec:LHC} from high-$p_T$ data with the ones obtained from flavor-physics observables, as discussed in Sec.~\ref{sec:lowenergy}. This comparison is made in Fig.~\ref{fig:EFT-comparison} where we depict the current and projected LHC limits from Table~\ref{tab:LHC-limits} at $36~\mathrm{fb}^{-1}$ (blue) and $3~\mathrm{ab}^{-1}$ (red), respectively. In the same plot, we include flavor constraints from quarkonium decays (light blue), $\mu\to e$ nuclear conversion (magenta), FCNC meson decays (green) and LFV $\tau$-decays (yellow). There are several interesting features of this plot which we discuss in the following.

Firstly, the high-$p_T$ limits on quark-flavor conserving Wilson coefficients $C_{q_iq_i}^{\ell_k \ell_l}$ are significantly better than the limits coming from quarkonium decays irrespective of the LFV dilepton pair. The latter are less competitive because they are obtained from measuring relatively wide (short lived) $q\bar{q}$ vector mesons ($\phi, J/\psi,\Upsilon$). Due to the large widths of these quarkonia, their LFV branching ratios are suppressed, and thus the low-energy bounds on the relevant Wilson coefficients are weaker. As a striking example, the high-$p_T$ bound on $C_{cc}^{\mu\tau}$ ($C_{ss}^{e\mu}$) is a factor of $\sim 300$ ($\sim 4\times 10^4$) times more stringent than the flavor bound. This conclusion can be extended to lepton flavor conserving transitions where analogous LHC searches in high-$p_T$ dilepton tails $q_i\bar q_i\to \ell^+\ell^-$ are expected to provide stronger bounds than the ones extracted from quarkonium decays.

\begin{table}[!t]
\centering
  \renewcommand{\arraystretch}{1.7} 
\begin{tabular}{|c|cc|}
\hline
\multicolumn{3}{|c|}{Selected LHC limits (left-handed scenario)} \\ \hline\hline
Decay mode & Current ($36~\mathrm{fb}^{-1}$) & Future ($3~\mathrm{ab}^{-1}$) \\
\hline\hline
\rowcolor{LightCyan}
$\phi\to \mu^\pm e^\mp$	& $ 8.7  \times 10^{-18} $ &  $ 1.2  \times 10^{-18} $  \\
\rowcolor{LightCyan}
$D^0\to \mu^\pm e^\mp$ & $ 3.1 \times 10^{-9}$ & $ 3.8 \times 10^{-10}$ \\
\rowcolor{LightCyan}
$J/\psi \to \mu^\pm e^\mp$	& $ 1.0 \times 10^{-11}$ & $ 1.6 \times 10^{-12} $ \\

$B_d\to \mu^\mp e^\pm$	& $ 9.7 \times 10^{-8}$	& $ 1.2 \times 10^{-8} $\\

$B^+\to \pi^+ \mu^\mp e^\pm $	& $  4.3  \times 10^{-5}$ & $ 5.6 \times 10^{-6} $  \\

$B_s\to \mu^\mp e^\pm$	& $ 9.1 \times 10^{-7}$  & $ 1.3 \times 10^{-7}$ \\

$B^+\to K^+ \mu^\mp e^\pm$ & $ 4.0  \times 10^{-4}$ & $ 5.6  \times 10^{-5}$ \\

$B^0\to K^\ast \mu^\mp e^\pm$ & $ 8.2 \times 10^{-4}$ & $  1.1 \times 10^{-4}$	\\ 

\rowcolor{LightCyan}
$ \Upsilon(3S) \to \mu^\pm e^\mp $	& $ 9.3  \times 10^{-9}$ &  $ 1.3 \times 10^{-9}$	\\

\hline \hline

\rowcolor{LightCyan}
$D^0 \to \tau^\pm e^\mp $	& $ 6.4 \times 10^{-8}$	& $ 2.7 \times 10^{-9}$	\\

\rowcolor{LightCyan}
$J/ \psi \to \tau^\pm e^\mp $	& $ 6.4 \times 10^{-11}$	& $ 4.8 \times 10^{-12}$	\\

\rowcolor{LightCyan}
$B_d\to \tau^\pm e^\mp$	& $ 2.0 \times 10^{-4}$	& $  2.0 \times 10^{-5}$	\\ 

\rowcolor{LightCyan}
$B^+ \to \pi^+ \tau^\pm e^\mp$	& $ 2.6 \times 10^{-4}$  &  $ 2.7 \times 10^{-5}$ \\ 

$B_s\to \tau^\mp e^\pm$	& $ 2.4 \times 10^{-3}$	& $ 1.7 \times 10^{-4}$ \\

$B^+\to K^+ \tau^\pm e^\mp$ & $ 3.1 \times 10^{-3}$  &  $ 2.1  \times 10^{-4}$	\\ 

$B^0\to K^\ast \tau^\pm e^\mp$ & $  5.1 \times 10^{-3}$ &  $  3.6 \times 10^{-4}$ \\

\rowcolor{LightCyan}
$ \Upsilon(3S) \to \tau^\pm e^\mp $	& $ 9.5 \times 10^{-8}$ &  $ 7.9 \times 10^{-9}$	\\

\hline\hline
\rowcolor{LightCyan}
$ J/\psi \to \tau^\pm \mu^\mp $	  & $ 6.8 \times 10^{-11}$ & $ 6.4 \times 10^{-12}$ \\

\rowcolor{LightCyan}
$B_d\to \tau^\pm \mu^\mp$	& $ 2.4 \times 10^{-4}$ & $ 1.8  \times 10^{-5}$ \\ 

\rowcolor{LightCyan}
$B^+ \to \pi^+ \tau^\pm \mu^\mp$	& $ 3.1 \times 10^{-4}$ & $  2.4 \times 10^{-5}$	\\ 

$B_s\to \tau^\pm \mu^\mp$	& $ 2.9 \times 10^{-3}$ & $ 2.5 \times 10^{-4}$	\\ 

$B^+\to K^+ \tau^\pm \mu^\mp$ & $ 3.5 \times 10^{-3}$ & $ 3.1 \times 10^{-4}$ \\

$B^0\to K^\ast \tau^\pm \mu^\mp$ & $  6.0 \times 10^{-3}$ & $ 5.3 \times 10^{-4}$ \\

\rowcolor{LightCyan}
$ \Upsilon(3S) \to \tau^\pm \mu^\mp $	& $ 1.0 \times 10^{-7}$	& $ 1.2 \times 10^{-8}$ \\

\hline
\end{tabular}
\caption{\em \small Limits on LFV branching fractions at $95\%$ obtained from the recast of high-$p_T$ dilepton tails in Table~\ref{tab:exp} for the left-handed scenario (cf.~Eq.~\eqref{eq:eft-simplified}). Decay modes for which the projected high-luminosity LHC limits are more stringent or comparable to the direct flavor ones (see Table~\ref{tab:exp}) are highlighted in blue. }   
\label{tab:exp-lhc}
\end{table}

Secondly, the low-energy limits from FCNC meson decays involving down-quarks or those from LFV $\tau$-decays are typically more constraining than high-$p_T$ dilepton tails at current luminosities. However, for some specific transitions, the constraints that we estimate at the high-luminosity phase of the LHC can become competitive with the limits derived from low-energy data. This is the case, for instance, in the tauonic channels where the HL-LHC bounds at $3~\mathrm{ab}^{-1}$ from $b\bar{d}\to e\tau$ and $b\bar{d}\to \mu\tau$ production become comparable to the LFV bounds from semi-leptonic $B\to \pi l\tau$ decay (with $l=e,\mu$). Note that a similar result was obtained in Ref.~\cite{Greljo:2018tzh} for the corresponding semi-tauonic charged current transition $b\to u \tau \bar{\nu}$, for which mono-tau production at the LHC can provide competitive limits with the current  limits on $B^0\to\pi^-\tau^+\nu$ from $B$-factories.

Another case for which LHC data provides meaningful constraints is the $c\to u$ transition. Interestingly, the only direct bound on the $C_{uc}^{e\tau}$ and $C_{uc}^{\mu\tau}$ Wilson coefficients comes from high-$p_T$ measurements. The corresponding low-energy two-body decays $D^0 \to \mu \tau$ and $\tau \to D^0 l$ (with $l=e,\mu$) are kinematically forbidden since $m_\tau>m_{D^0}$ and $\left| m_{D^0} - m_{\tau} \right| < m_{\mu}$, whereas the $D^0 \to e \tau$ decay is strongly phase-space suppressed (see also Ref.~\cite{Dorsner:2013tla}). This result also extends to lepton flavor conserving transitions involving charm quarks where the LHC dilepton tails are expected to provide competitive limits in comparison to (semi)leptonic $D$-meson decays, cf.~Ref.~\cite{Greljo:2017vvb}.

To make the complementarity between flavor physics and LHC constraints even more explicit, we translate the LHC limits obtained in Table~\ref{tab:LHC-limits} into bounds on the corresponding LFV decays,  for the benchmark scenario with purely left-handed defined in Eq.~\eqref{eq:eft-simplified}, as shown in Table~\ref{tab:exp-lhc}. We do not consider the processes that involving the unflavored mesons $\pi^0$, $\omega$, $\rho$ and $K_L$, since they would depend on several Wilson coefficients, making this comparison less straightforward, cf.~e.g.~Eq.~\eqref{eq:eqEx1}--\eqref{eq:eqEx3}. For the remaining processes, we obtain indirect limits on the branching fractions which can be directly compared to Table~\ref{tab:exp}, reinforcing the conclusions drawn above. For instance, an improvement of the experimental sensitivity on the quarkonium decay rates by several orders of magnitude would be needed to make them comparable to the LHC constraints, as discussed above. For $B$-meson decays, there is an interplay between low and high-energy constraints, as one can see for example by comparing the current experimental limit on $\mathcal{B}(B\to\pi \mu^\pm \tau^\mp)^\mathrm{exp}<9.4\times 10^{-5}$ (95\%CL)~\cite{Tanabashi:2018oca}, with the projected limit for the LHC high-luminosity phase that we obtain, namely $\mathcal{B}(B\to\pi \mu^\pm \tau^\mp)\lesssim 2.4\times 10^{-5}$. Lastly, we are able to obtain the indirect limit $\mathcal{B}(D^0\to e \tau)< 2.7\times 10^{-8}$ ($95\%$ CL), for which there is  no experimental search yet. Note that these conclusions are only valid for scenarios based on left-handed operators (cf.~Eq.~\ref{eq:eft-simplified}). The relative importance of direct flavor constraints and the indirects ones inferred from high-$p_T$ data will certanly change if scalar and tensor operators are also present. In this paper, we do not perform a comparison between flavor and LHC constraints for the most general new physics scenario that include these operators, but we provide all the needed ingredients for such analysis in Appendix~\ref{app:flavor}.

Finally, we comment on the validity of the EFT formulation when quoting high-$p_T$ bounds on LFV Wilson coefficients. In our definition from Eq.~\eqref{eq:eft-simplified}, we absorb a factor of $v^2/\Lambda^2$ into the Wilson coefficients, where the cutoff $\Lambda$ corresponds approximately to the mass of a heavy mediator which has been integrated out at tree level. For left-handed operators there are two possible ultra-violet completions one can consider: (i) a color-singlet vector boson, i.e.~a $Z^\prime$, or (ii) a color-triplet vector boson, i.e.~a leptoquark. These particles would contribute to dilepton production via the $s$- and $t$-channel, respectively. In both cases, if the mass of the new mediator is lower than the energy scale involved in this process, the EFT expansion would breakdown and our high-$p_T$ EFT limits cannot be used. This breakdown is not so significant for $t$-channel mediators, since the cross-sections computed in the EFT and full model do not differ significantly~\cite{Davidson:2013fxa}. However, this reinterpretation of EFT constraints would be very problematic for $s$-channel mediators such as $Z^{\prime}$ bosons. To put this on more quantitative grounds, we studied the applicability of our EFT limits by directly comparing the bounds of the EFT with those obtained from a concrete LFV $Z^{\prime}$ model with couplings to bottom quarks. We found that the limits extracted from dilepton tails in the $Z^\prime$ coupling--mass $(g_\ast,M_{Z^\prime})$ plane quickly converge to the EFT limits for mediator masses $M_{Z^\prime}$ above the $4-6$ TeV range. Below this mass range the limits quoted in Table~\ref{tab:LHC-limits} are not valid anymore and the limits extracted from the complete model should be used. 

\section{Summary}
\label{sec:conclusion}

In this paper, we have derived limits on LFV quark-lepton dimension-6 operators by using LHC data from $pp \to \ell_i \ell_j$ tails (with $i \neq j$) at high-$p_T$. For left-handed operators, these limits are summarized in Table~\ref{tab:LHC-limits}, which represents the central result of this paper. For the case of general semi-leptonic operators, including the scalar and tensor ones, the results from Table~\ref{tab:LHC-limits} can be adapted by using the guidelines from Sec.~\ref{ssec:summary} and, in particular, Eq.~\eqref{eq:stat}.

For the specific case of left-handed semi-leptonic operators, we have compared the bounds coming from dilepton tails with the low-energy flavor bounds, highlighting the complementarity between the two approaches. We have found that, in the case of operators violating quark flavor as well, low-energy constraints coming from FCNC meson or $\tau$-lepton decays provide in most cases much tighter bounds compared to the high-$p_T$ constraints. The only exception to this rule involves the $C_{uc}^{e\tau}$ and $C_{uc}^{\mu\tau}$ Wilson coefficients, which are not constrained at all by flavor measurements due to the fact that there is no experimental search for $D^0 \to e\tau$, which is heavily phase-space suppressed, while the $D^0 \to \mu\tau$ and $\tau\to\mu D^0$ decays are kinematically forbidden.

We have also found that operators that conserve quark flavor are generally better constrained by high-$p_T$ dilepton tails at the LHC. In particular, quarkonium decays provide relatively weak bounds on the effective coefficients, which are thus better constrained by the LHC dilepton tails. Notice that this result also extends to lepton flavor conserving operators, that is, LHC searches in the dilepton tails $q\bar q\to ee,\mu\mu,\tau\tau$ provide much stronger limits than the corresponding quarkonium decay limits from low energy experiments.  As exceptions, $\mu \to e$ conversion in nuclei and LFV $\tau$ decays involving light unflavored mesons such as $\tau \to \ell \rho$ and $\tau \to \ell \phi$ provide more competitive bounds on the relevant Wilson coefficients compared to high-$p_T$ dilepton production. Another interesting example are the decays $B\to \pi \tau l$, with $l=e,\mu$, for which the projected HL-LHC limits become more constraining than the present flavor limits. All these comparisons are summarized in Fig.~\ref{fig:EFT-comparison}. Finally, to further illustrate the complementarity of both approaches for the benchmark scenario with purely left-handed operators, we translate the high-$p_T$ bounds from Table~\ref{tab:LHC-limits} into limits on the corresponding low-energy processes, as shown in Table~\ref{tab:exp-lhc}. Decay modes for which LHC constraints in the high-luminosity phase will be more stringent than low-energy constraints are highlighted in blue, reinforcing the conclusions drawn above.

\section{Acknowledgments}
\label{sec:acknowledgments}

We thank D.~Be\v{c}irevi\'{c} for numerous stimulating discussions and for encouraging us to pursue this project. We also thank G.~Isidori, S.~Fajfer, J.~Fuentes-Martín, C.~Joo, T.~Kitahara, F.~Mescia and P.~Paradisi for helpful discussions. This project has received support by the European Union's Horizon 2020 research and innovation programme under the Marie Sklodowska-Curie grant agreement N$^\circ$~674896. A.A. acknowledges support from University of Nebraska-Lincoln, National Science Foundation under grant number PHY-1820891, and the NSF Nebraska EPSCoR under grant number OIA-1557417.~D.A.F. is supported by the Swiss National Science Foundation (SNF) under contract
200021-159720.

\appendix

\section{Matching to the Warsaw basis}
\label{app:warsaw}

In this Appendix we provide the tree-level matching of Eq.~\eqref{eq:left} to the Warsaw basis. We consider the same notation for the operators of Ref.~\cite{Jenkins:2013zja,Jenkins:2013wua,Alonso:2013hga} and we assume that down-quark Yukawas are diagonal. Operators with down and up-type quarks are treated separately as they can arise from different operators in the SMEFT:

For \emph{down-type quark} operators in Eq.~\eqref{eq:left}, we find 
\begin{align}
\label{eq:cvll-d}
C_{V_{LL}}^{ijkl} &=\dfrac{v^2}{\Lambda^2} \,\bigg{(}C_{\substack{lq\\ klij}}^{(1)} + C_{\substack{lq\\ klij}}^{(3)} \bigg{)}\,,\\
C_{V_{RR}}^{ijkl} &=\dfrac{v^2}{\Lambda^2} \,C_{\substack{ed\\klij}}\,,\\
\label{eq:cvlr-d}
C_{V_{LR}}^{ijkl} &=\dfrac{v^2}{\Lambda^2} \,C_{\substack{qe\\ijkl}}\,,\\
C_{V_{RL}}^{ijkl} &=\dfrac{v^2}{\Lambda^2} \,C_{\substack{ld\\klij}}\,,\\
C_{S_{R}}^{ijkl} &=\dfrac{v^2}{\Lambda^2} \,C_{\substack{ledq\\klij}}\,,\\
\label{eq:cst-d}
C_{S_{L}}^{ijkl} &= C_{T}^{ijkl} =0\,.
\end{align}

 For \emph{up-type quarks} operators,
\begin{align}
\label{eq:cvll-u}
C_{V_{LL}}^{ijkl} &= \dfrac{v^2}{\Lambda^2}\,V_{ip}V_{jr}^\ast\,\bigg{(}C_{\substack{lq\\ klpr}}^{(1)} - C_{\substack{lq\\ klpr}}^{(3)} \bigg{)}\,,\\
\label{eq:cvlr-u}
C_{V_{RR}}^{ijkl} &=\dfrac{v^2}{\Lambda^2} \,C_{\substack{eu\\klij}}\\
C_{V_{LR}}^{ijkl} &=\dfrac{v^2}{\Lambda^2}
\,V_{ip}V_{jr}^\ast\, C_{\substack{qe\\prkl}}\,,\\
C_{V_{RL}}^{ijkl} &= \dfrac{v^2}{\Lambda^2}
\, C_{\substack{lu\\klij}}\,,\\
\label{eq:csl-u}
C_{S_{L}}^{ijkl} &=- \dfrac{v^2}{\Lambda^2}\,V_{ip}\,\,C_{\substack{lequ\\klpj}}^{(1)}\,,\\
\label{eq:ct-u}
C_{T}^{ijkl} &=-\dfrac{v^2}{\Lambda^2}\,V_{ip}\,C_{\substack{lequ\\klpj}}^{(3)}\,,\\
\label{eq:csr-u}
C_{S_{R}}^{ijkl} &= 0\,,
\end{align}

\noindent where $V\equiv V_{\mathrm{CKM}}$ denotes the CKM matrix and the summation over repeated flavor indices is implicit. Right-handed fermions are assumed to be in the mass basis. Contributions induced by renormalization group evolution are neglected in the above equations. 

The equations given above can now be combined with Eq.~\eqref{eq:stat} to constrain any effective scenario formulated above the electroweak scale. We stress once again that both up and down-type quark flavors should be added in Eq.~\eqref{eq:stat}, since they can both contribute to the cross-section. For operators involving quark doublets, one should be careful as different quark-flavor combinations are induced by the CKM matrix, which should then be added in Eq.~\eqref{eq:stat}, cf.~Eqs.~\eqref{eq:cvll-u}, \eqref{eq:cvlr-u}, \eqref{eq:csl-u} and \eqref{eq:ct-u}.

\section{General expressions for meson decays}
\label{app:flavor}

In this Appendix we generalize the expressions for LFV meson decays, accounting for all operators introduced in Eq.~\eqref{eq:left}. In the following, we consider decays based on the transition $q_{j}\to q_{i} \ell_{k}^- \ell_{l}^+$, with $k>l$. To express the decay rates in a compact form, it is convenient to consider operators with a definite parity in the quark current,~\footnote{These expressions should be corrected for processes involving neutral kaons, as will be discussed in the following.}
\begin{align}
\label{eq:wc-simple}
\mathscr{C}_{(^{S}_{P})R} &= \dfrac{C_{S_{L}}^{ijkl} \pm C_{S_{R}}^{ijkl}}{2}\,, \qquad\mathscr{C}_{(^{V}_{A})X} = \dfrac{C_{V_{RX}}^{ijkl} \pm C_{V_{LX}}^{ijkl}}{2}\,, \nonumber\\[0.3em]
\mathscr{C}_{(^{S}_{P})L} &= \dfrac{\big{(}C_{S_{R}}^{jilk}\big{)}^\ast\pm \big{(}C_{S_{L}}^{jilk}\big{)}^\ast}{2}\,,
\end{align}

\begin{table*}[!t]
\renewcommand{\arraystretch}{1.9}
\centering
\begin{tabular}{|c|cccccccccccc|}
\hline 
$P\to M \ell_i\ell_j$ & $a_{V}^{+}$ & $a_{V}^{-}$ & $a_{A}^{+}$ & $a_{A}^{-}$ & $a_S^{+}$ & $a_S^{-}$ & $a_P^{+}$ & $a_P^{-}$ & $a_{VS}^{+}$ & $a_{VS}^{-}$ & $c_{AP}^{+}$ & $c_{AP}^{-}$ \\ \hline\hline
$K^+\to \pi^+ e^+ \mu^-$ & $0.596(4)$ & $0.598(4)$ & $0$ & $0$ & $9.79(6)$ & $9.84(6)$ & $0$ & $0$ & $2.70(2)$ &  $2.74(2)$ & $0$ & $0$ \\ 
$K_L\to \pi^0 e^+ \mu^-$ & $2.75(2)$ & $2.76(2)$ & $0$ & $0$ & $47.5(3)$ & $47.7(3)$ & $0$ & $0$ & $12.6(1)$ &  $12.8(1)$ & $0$ & $0$ \\ 
$K_S\to \pi^0 e^+ \mu^-$ & $0.00480(4)$ & $0.00482(4)$ & $0$ & $0$ & $0.0831(6)$ & $0.0834(6)$ & $0$ & $0$ & $0.0220(2)$ &  $0.0223(2)$ & $0$ & $0$ \\ \hline\hline
$B\to \pi e^+ \mu^-$ & $5.7(4)$ & $5.7(4)$ & $0$ & $0$ & $8.1(5)$ & $8.1(5)$ & $0$ & $0$ & $0.50(3)$ & $0.50(3)$ & $0$ & $0$ \\
$B\to \pi e^+ \tau^-$ & $3.7(2)$ & $3.7(2)$ & $0$ & $0$ & $5.2(3)$ & $5.2(3)$ & $0$ & $0$ & $4.2(3)$ & $4.2(3)$ & $0$ & $0$\\
$B\to \pi \mu^+\tau^-$ & $3.6(2)$ & $3.7(2)$ & $0$ & $0$ & $5.0(3)$ & $5.3(3)$ & $0$ & $0$ & $3.8(2)$ & $4.6(3)$ & $0$ & $0$\\ \hline\hline 
$B\to K e^+ \mu^-$ & $8.2(6)$ & $8.2(6)$ & $0$ & $0$ & $14.5(6)$ & $14.5(6)$ & $0$ & $0$  & $1.07(7)$ & $1.09(7)$ & $0$ & $0$\\
$B\to K e^+ \tau^-$ & $5.3(2)$ & $5.3(2)$ & $0$ & $0$ & $8.4(3)$ & $8.4(3)$ & $0$ & $0$ & $8.1(3)$ & $8.1(3)$ & $0$ & $0$\\
$B\to K \mu^+\tau^-$ & $5.2(2)$ & $5.2(2)$ & $0$ & $0$ & $8.1(2)$ & $8.7(3)$  & $0$ & $0$ & $7.3(3)$ & $8.9(4)$ & $0$ & $0$\\ \hline\hline
$B\to K^\ast  e^+ \mu^-$ & $2.8(5)$ & $2.8(5)$ & $14(2)$ & $14(2)$ &  $0$& $0$ & $4.6(8)$ & $4.6(8)$ & $0$ & $0$ & $-0.5(1)$ & $-0.5(1)$ \\
$B\to K^\ast e^+ \tau^-$ & $1.4(2)$ & $1.4(2)$ & $7.5(8)$ & $7.5(8)$ & $0$ & $0$ & $2.0(3)$ & $2.0(3)$ & $0$ & $0$ & $-2.6(5)$ & $-2.6(5)$ \\
$B\to K^\ast \mu^+\tau^-$ & $1.5(2)$ & $1.3(2)$ & $7.6(8)$ & $7.1(8)$ & $0$ & $0$ & $1.9(3)$ & $2.1(4)$ & $0$ & $0$ & $-2.3(4)$ & $-2.9(5)$ \\ \hline 
\end{tabular}
\caption{ \sl Values for the multiplicative factors defined in Eq.~\eqref{eq:semi-leptonic-formulas} computed by using the form factor for the transitions $K\to \pi$, $B\to \pi$, $B\to K$ and $B\to K^\ast$ reported in Ref.~\cite{Carrasco:2016kpy}, \cite{Lattice:2015tia}, \cite{Aoki:2019cca} and \cite{Straub:2015ica}, respectively. The effective coefficients to be replaced in Eq.~\eqref{eq:semi-leptonic-formulas} are defined in Eq.~\eqref{eq:wc-simple} for $B$-meson and $K^+$ decays, and in Eq.~\eqref{eq:kaons-cv-cs} for $K_{L(S)}$ decays.}
\label{tab:LFV-formulas} 
\end{table*}

\noindent where $X=L,R$, as before, and the upper (lower) subscript correspond to a plus (minus) sign in the expressions. We also define $\mathscr{C}_T = C_T^{ijkl}$ and $\widehat{\mathscr{C}}_T = \big{(}C_T^{jilk}\big{)}^\ast$. We assume that these Wilson coefficients are evaluated at the same scale $\mu$ in which the hadronic parameters have been determined. For scalar and tensor operators, the QCD running from $\Lambda \approx 1$~TeV down to $m_b$ is known to be sizeable, see Ref.~\cite{Gracey:2000am} and references therein. Furthermore, the electroweak running can induce a non-negligible mixing of $\mathcal{O}_{T}$ into $\mathcal{O}_{S_L}$~\cite{Gonzalez-Alonso:2017iyc,Feruglio:2018fxo}.

\

\paragraph*{$\bullet$~\underline{$P\to \ell_k \ell_l$}} We first consider the leptonic decays of a pseudoscalar meson of type $P=\bar{q}_i q_j$, for which it is straightforward to show that~\cite{Becirevic:2016zri}
\begin{align}
\label{eq:Pllbis}
\mathcal{B}(P\to &\ell_k^- \ell_l^+) =  \dfrac{\tau_P\,f_P^2 m_P m_{\ell_k}^2}{16 \pi v^4}\left(1-\dfrac{m_{\ell_k}^2}{m_P^2}\right)^2\\ 
&\times\Bigg{\lbrace}\bigg{|}\mathscr{C}_{AL}- \dfrac{\mathscr{C}_{PL}\,m_{P}^2}{m_{\ell_k} (m_{q_i}+m_{q_j})}\bigg{|}^2 +  (L \leftrightarrow R)\Bigg{\rbrace}\,,\nonumber
\end{align}
where we have used $m_{\ell_k}\gg m_{\ell_l}$ to simplify the expression, and the decay constant $f_P$ is defined in the usual way, namely $\langle 0 | \bar{q}_i \gamma^\mu \gamma_5 q_j | P(p) \rangle = i \,f_P \, p^\mu$. The most recent determination of $f_P$ for the relevant mesons are summarized in Ref.~\cite{Aoki:2019cca}. Note that Eq.~\eqref{eq:Pllbis} should be amended for the neutral kaon system, $K_{L(S)}\simeq (K^0 \pm \overbar{K^0})/\sqrt{2}$, by making the following replacements,
\begin{align}
\label{eq:kaons-ca-cp}
\begin{split}
\mathscr{C}_{AX}& \to \dfrac{C_{V_{RX}}^{sdkl} - C_{V_{LX}}^{sdkl}}{2\sqrt{2}} \pm (s \leftrightarrow d) \,, \\[0.3em]
\mathscr{C}_{PR}& \to \dfrac{C_{S_{L}}^{sdkl} - C_{S_{R}}^{sdkl}}{2\sqrt{2}}\pm(s \leftrightarrow d) \\[0.3em]
\mathscr{C}_{PL}& \to \dfrac{\big{(}C_{S_{R}}^{dslk}\big{)}^\ast - \big{(}C_{S_{L}}^{dslk}\big{)}^\ast}{2\sqrt{2}}\pm(s \leftrightarrow d)\,,
\end{split}
\end{align}
where the plus (minus) sign corresponds to $K_L$ ($K_S$). Furthermore, note that the expression for the mode with conjugate electric charge, i.e.~$\mathcal{B}(P\to \ell_k^+ \ell_l^-)$, is analogous to Eq.~\eqref{eq:Pllbis} with the replacement $\mathscr{C}_{PL}\leftrightarrow - \mathscr{C}_{PR}$, which can be understood from the non-conservation of the LFV vector current,~$i\partial_\mu(\bar{\ell}_k \gamma^\mu \ell_l)=(m_{\ell_l}-m_{\ell_k}) \,\bar{\ell}_k \ell_l$.

\

\paragraph*{$\bullet$~\underline{$\tau\to \ell_l P$}} The general expression for $\tau \to \ell_l P$ decays (with $l=e,\mu$) reads
\begin{align}
\mathcal{B}(\tau^-\to &\ell_l^- P) = \dfrac{\tau_\tau\, f_P^2 m_\tau^3}{32\pi v^4}\left(1- \dfrac{m_P^2}{m_\tau^2} \right)^2\\ 
&\times\Bigg{\lbrace}\bigg{|}\mathscr{C}_{AL}+ \dfrac{\mathscr{C}_{PR}\,m_{P}^2}{m_{\ell_k} (m_{q_i}+m_{q_j})}\bigg{|}^2 +  (L \leftrightarrow R)\Bigg{\rbrace}\,,\nonumber
\end{align}
\noindent where lepton flavor indices in Eq.~\eqref{eq:wc-simple} are such that $k=\tau$. For decays into kaons, one should use the replacements given in Eq.~\eqref{eq:kaons-ca-cp}. Similarly to previous observable, the expression for the decay with opposite leptonic charges can be obtained by replacing $\mathscr{C}_{PR}\leftrightarrow - \mathscr{C}_{PL}$. 

\

\paragraph*{$\bullet$~\underline{$V\to \ell_k \ell_l$}} Next, we consider leptonic decays of quarkonia $V \in \lbrace \psi,J/\psi, \Upsilon\rbrace$. We obtain
\begin{align}
\mathcal{B}(V &\to \ell_k^- \ell_l^+) = \dfrac{\tau_V\, m_V^3 f_V^2}{24\pi v^4}\bigg{(}1-\dfrac{m_{\ell_k}^2}{m_V^2}\bigg{)}^2\nonumber\\
&\times\Bigg{\lbrace} \bigg{[}|\mathscr{C}_{VL}|^2 + |\mathscr{C}_{VR}|^2\bigg{]} \bigg{(}1+\dfrac{m_{\ell_k}^2}{2m_V^2}\bigg{)} \\
&\quad+6\dfrac{f_V^T}{f_V}\dfrac{m_{\ell_k}}{m_V} \mathrm{Re}\big{(}\mathscr{C}_T\, \mathscr{C}_{VR}^\ast+\widehat{\mathscr{C}}_T\, \mathscr{C}_{VL}^\ast \big{)} \,, \nonumber\\
&\quad+ 2\bigg{(}\dfrac{f_V^T}{f_V}\bigg{)}^2 \bigg{[}|\mathscr{C}_{T}|^2 + |\widehat{\mathscr{C}}_{T}|^2\bigg{]}\bigg{(}1+\dfrac{2 m_{\ell_k}^2}{m_V^2}\bigg{)}\nonumber\Bigg{\rbrace}\,,
\end{align}
where $f_V$ and $f_V^T$ stands for the vector and tensor decay constants, which are defined by
\begin{align}
\begin{split}
\langle 0 | \bar{q}\gamma^\mu q | V(p,\lambda) \rangle&= f_V m_V e^\mu(\lambda)\,,\\[0.3em]
\langle 0 | \bar{q}\sigma_{\mu\nu}q | V(p,\lambda) \rangle &= i f_V^T \left[e_\mu(\lambda) p_\nu - e_\nu(\lambda) p_\mu\right]\,,
\end{split}
\end{align}
where $e_\mu(\lambda)$ is the polarization vector of $V= q\bar{q}$. See Ref.~\cite{Donald:2013pea}, \cite{Becirevic:2012dc} and \cite{Colquhoun:2014ica,Lewis:2012ir} for recent lattice QCD determinations for $\phi$, $J/\psi$ and $\Upsilon$, respectively. The tensor decay constant has also been computed on the lattice for $J/\psi$ and it is found to be similar to the vector one, i.e.~$f_\psi^T \approx f_\psi$~\cite{Becirevic:2012dc}. Note also that the interference term in the above expression changes sign for the charge-conjugate mode. 

\

\paragraph*{$\bullet$~\underline{$\tau\to \ell_l V$}} We also compute the expressions for LFV $\tau$ decays into vector mesons such as $V=K^\ast, \rho$ or $\phi$, for which we find
\begin{align}
\mathcal{B}(\tau \to \ell_l V) &= \dfrac{\tau_\tau\, m_\tau^3 f_V^2}{32\pi v^4}\bigg{(}1-\dfrac{m_{V}^2}{m_{\tau}^2}\bigg{)}^2\nonumber\\
&\times\Bigg{\lbrace} \bigg{[}|\mathscr{C}_{VL}|^2 + |\mathscr{C}_{VR}|^2\bigg{]} \bigg{(}1+\dfrac{2 m_{V}^2}{m_{\tau}^2}\bigg{)} \\
&\quad+12\dfrac{f_V^T}{f_V}\dfrac{m_{V}}{m_\tau}\mathrm{Re}\big{(}\mathscr{C}_T\, \mathscr{C}_{VL}^\ast+\widehat{\mathscr{C}}_T\, \mathscr{C}_{VR}^\ast \big{)} \,, \nonumber\\
&\quad+ 8\bigg{(}\dfrac{f_V^T}{f_V}\bigg{)}^2 \bigg{[}|\mathscr{C}_{T}|^2 + |\widehat{\mathscr{C}}_{T}|^2\bigg{]}\bigg{(}1+\dfrac{2 m_{V}^2}{2m_\tau^2}\bigg{)}\nonumber\Bigg{\rbrace}\,,
\end{align}

\noindent where leptonic flavor indices in Eq.~\eqref{eq:wc-simple} are such that $k=\tau$. The vector (tensor) decay constants can be found in Ref.~\cite{Allton:2008pn}.

\

\paragraph*{$\bullet$~\underline{$P\to M \ell_k \ell_l $}} Lastly, we provide general expression for the most relevant semi-leptonic decays, namely the one based on the transitions $K\to \pi$, $B\to \pi$ and $B\to K^{(\ast)}$. We focus on vector and scalar operators, since tensor operators are absent at dimension-6 in the SMEFT for $d_i\to d_j \ell_k^- \ell_l^+$ decays, cf.~Eq.~\eqref{eq:cst-d}. We parametrize the general branching fractions as~\cite{Becirevic:2016zri}
\begin{align}
\label{eq:semi-leptonic-formulas}
\mathcal{B}(P \to M \ell_k^- \ell_l^+) &= \sum_{\alpha}\big{[}a_\alpha^{+}\, |\mathscr{C}_{\alpha,L+R}|^2+a_\alpha^{-}\, |\mathscr{C}_{\alpha,L-R}|^2\big{]}\nonumber\\
&+a_{VS}^{+}\,\mathrm{Re}[\mathscr{C}_{V,L+R}\,(\mathscr{C}_{S,L+R})^\ast]\nonumber\\[0.3em]
&+a_{VS}^{-}\,\mathrm{Re}[\mathscr{C}_{V,L-R}\,(\mathscr{C}_{S,L-R})^\ast]\nonumber\\[0.3em]
&+a_{AP}^{+}\,\mathrm{Re}[\mathscr{C}_{A,L+R}\,(\mathscr{C}_{P,L+R})^\ast]\nonumber\\[0.3em]
&+a_{AP}^{-}\,\mathrm{Re}[\mathscr{C}_{A,L-R}\,(\mathscr{C}_{P,L-R})^\ast]\,,
\end{align}

\noindent where the summation extends over $\alpha=\lbrace V,S,P,A\rbrace$, and $M$ denotes a generic pseudoscalar or vector meson. The effective coefficients are defined in Eq.~\eqref{eq:wc-simple} for charged kaon and $B$-meson decays, which are evaluated at $\mu=2$~GeV and $\mu=m_b$, respectively. For neutral kaons, one should use instead

\begin{align}
\label{eq:kaons-cv-cs}
\begin{split}
\mathscr{C}_{VX}& \to \dfrac{C_{V_{RX}}^{sdkl} + C_{V_{LX}}^{sdkl}}{4 } \mp (s \leftrightarrow d) \,, \\[0.3em]
\mathscr{C}_{SR}& \to \dfrac{C_{S_{L}}^{sdkl} + C_{S_{R}}^{sdkl}}{4}\pm(s \leftrightarrow d) \\[0.3em]
\mathscr{C}_{SL}& \to \dfrac{\big{(}C_{S_{R}}^{dslk}\big{)}^\ast + \big{(}C_{S_{L}}^{dslk}\big{)}^\ast}{4}\pm(s \leftrightarrow d)\,,
\end{split}
\end{align}

\vspace*{0.4em}

\noindent where the upper (lower) sign corresponds to the $K_L\to\pi^0 \mu e$ ($K_S\to\pi^0 \mu e$) decay, and $k,l \in \{e,\mu\}$. The values for the numeric coefficients $a_i^{\pm}$ are collected in Table~\ref{tab:LFV-formulas}. We have used the $K\to \pi$~\cite{Carrasco:2016kpy} and $B\to K$~\cite{Bouchard:2013pna,Bailey:2015dka} form factors computed on the lattice, see also~Ref.~\cite{Aoki:2019cca}. For the $B\to \pi$ transition, we have use the combined fit of experimental and LQCD data from Ref.~\cite{Lattice:2015tia}. For the $B\to K^\ast$ transition we use the results from Ref.~ \cite{Straub:2015ica}.

\end{document}